\documentclass[pdflatex,sn-basic]{sn-jnl}% Basic Springer Nature Reference Style/Chemistry Reference Style
\usepackage{graphicx}%
\usepackage{multirow}%
\usepackage{amsmath,amssymb,amsfonts}%
\usepackage{amsthm}%
\usepackage{mathrsfs}%
\usepackage[title]{appendix}%
\usepackage{xcolor}%
\usepackage{textcomp}%
\usepackage{manyfoot}%
\usepackage{booktabs}%
\usepackage{algorithm}%
\usepackage{algorithmicx}%
\usepackage{algpseudocode}%
\usepackage{listings}%
\usepackage{xspace}

\newcommand{\ourname}{MRI-CORE\xspace}

\newcommand{\ie}{\textit{i.e.}}

\usepackage[switch]{lineno}
\theoremstyle{thmstyleone}%
\theoremstyle{thmstyletwo}%

\theoremstyle{thmstylethree}%

\begin{document}

\title[\ourname]{\ourname: A Foundation Model for Magnetic Resonance Imaging}


\author*[1]{\fnm{Haoyu} \sur{Dong}}\email{haoyu.dong151@duke.edu}

\author[1]{\fnm{Yuwen} \sur{Chen}}\email{yuwen.chen@duke.edu}

\author[1]{\fnm{Hanxue} \sur{Gu}}\email{hanxue.gu@duke.edu}

\author[1]{\fnm{Nicholas} \sur{Konz}}\email{nicholas.konz@duke.edu}

\author[1]{\fnm{Yaqian} \sur{Chen}}\email{yaqian.chen@duke.edu}

\author[2]{\fnm{Qihang} \sur{Li}}\email{qihang.li@duke.edu}

\author[1,2,3,4]{\fnm{Maciej A.} \sur{Mazurowski}}\email{maciej.mazurowski@duke.edu}

\affil[1]{\orgdiv{Department of Electrical and Computer Engineering}, \orgname{Duke University}, \orgaddress{\street{2080 Duke University Road}, \city{Durham}, \postcode{27708}, \state{NC}, \country{USA}}}

\affil[2]{\orgdiv{Department of Biostatistics and Bioinformatics}, \orgname{Duke University}, \orgaddress{\street{2080 Duke University Road}, \city{Durham}, \postcode{27708}, \state{NC}, \country{USA}}}

\affil[3]{\orgdiv{Department of Radiology}, \orgname{Duke University}, \orgaddress{\street{2080 Duke University Road}, \city{Durham}, \postcode{27708}, \state{NC}, \country{USA}}}

\affil[3]{\orgdiv{Department of Computer Science}, \orgname{Duke University}, \orgaddress{\street{2080 Duke University Road}, \city{Durham}, \postcode{27708}, \state{NC}, \country{USA}}}

\abstract{
The widespread use of Magnetic Resonance Imaging (MRI) in combination with deep learning shows promise for many high-impact automated diagnostic and prognostic tools. However, training new models requires large amounts of labeled data, a challenge due to high cost of precise annotations and data privacy. To address this issue, we introduce the MRI-CORE, a vision foundation model trained using more than 6 million slices from over 110 thousand MRI volumes across 18 body locations. Our experiments show notable improvements in performance over state-of-the-art methods in 13 data-restricted segmentation tasks, as well as in image classification, and zero-shot segmentation, showing the strong potential of MRI-CORE to enable data-efficient development of artificial intelligence models. We also present data on which strategies yield most useful foundation models and a novel analysis relating similarity between pre-training and downstream task data with transfer learning performance. Our model is publicly available with a permissive license.
}




\maketitle

\section{Main}\label{sec1}

% mri important
Magnetic Resonance Imaging (MRI) is one of the most widely used imaging modalities in medical diagnostics, with around 100-150 million scans performed annually worldwide \citep{papanicolas2018health}. MRI supports a wide range of clinical tasks, including lesion detection, tissue classification, and disease monitoring. Among these tasks, segmentation plays a particularly important role, as it enables precise delineation of anatomical structures and pathological regions, directly impacting diagnosis, treatment planning, and longitudinal studies \citep{mazurowski2023segment,ma2024segment,azad2024medical,xu2024advances}. Recent advances in deep learning have significantly improved the automation and accuracy of MRI-based analyses across a variety of tasks. However, deep learning-based methods typically require large amounts of manually annotated data and lack task transferability, making them difficult to scale across new tasks, anatomies, or patient populations.

% no fm for mri + segmentation
Building large-scale pre-trained (typically in a self-supervised manner) models is a solution to these issues because of their potentially strong transfer learning and generalization capabilities. This approach has been demonstrated in general computer vision \citep{caron2021emerging, he2022masked,grill2020bootstrap} and pathology \citep{chen2024towards, lu2024visual, vorontsov2023virchow}. Inspired by these efforts, similar directions have emerged in radiology \citep{wu2024voco, li2024well, sun2025foundation}. However, these works usually utilize multiple image modalities, including CT, MRI, and X-Ray, during pre-training, which can dilute modality-specific representations, and often lacks comprehensive coverage of all anatomical regions. The most related study is \citep{wang2025triad}, yet its pre-training dataset only covered three body regions (brain, breast and prostate). Furthermore, this model was designed specifically for 3D inputs; 2D-based models remain dominant due to their lower computational requirements, often better performance (given the available training data), compatibility with slice-wise annotations, and ease of integration into existing workflows. We hypothesize that the development of a slice-based foundation model based on a wide range of MRI locations, such as the one proposed in this study, will strike a balance between well-founded representations based on millions of image examples and yet focus on relevant features, given its focus on MRIs. Since MRIs are one of the most common medical images and 2D analysis remains dominant, we also anticipate a wide applicability of our model across many real-world applications.

\begin{figure}[htbp]
    \centering
    \includegraphics[width=1\linewidth]{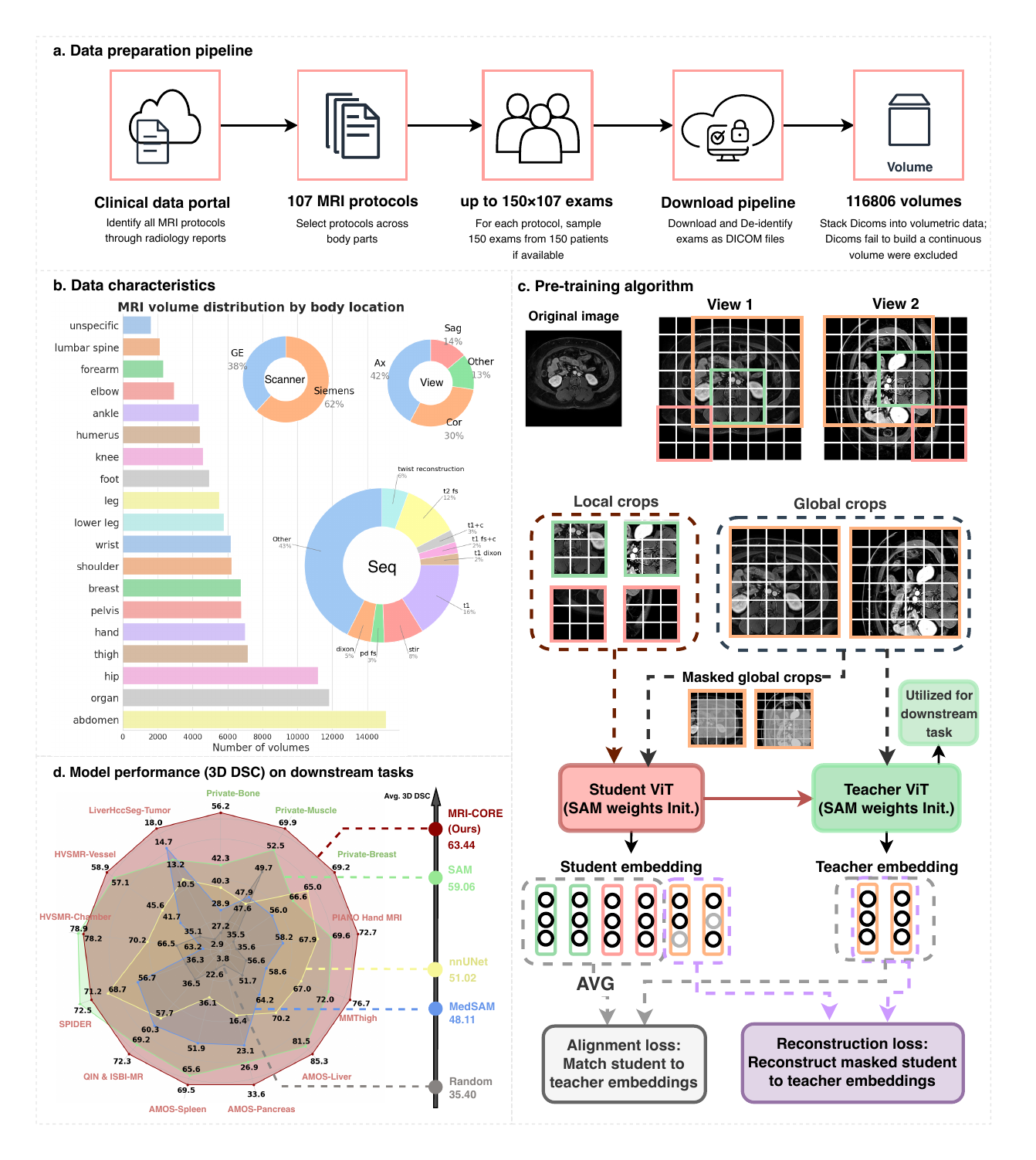}
    \caption{
    \textbf{a}. Data acquisition pipeline. First 107 protocols that cover different body parts were selected. For each protocol, we sampled up to 150 exams from different patients. All exams were de-identified and exported, resulting in 116,806 volumes.
    \textbf{b}. Distribution of the dataset in terms of body locations, scanner types, view, and sequence. 
    \textbf{c}. \ourname was pretrained using the DINOv2 self-supervised learning algorithm. Both the student and teacher ViT were initialized with SAM's weight, and the teacher ViT would be used as the final model.  
    \textbf{d}. Radar plot comparing the performance of \ourname and baselines on various segmentation tasks. \ourname outperformed both generalists (SAM, MedSAM) and specialists (nnU-Net) by a significant margin. 
    }
    \label{fig:summary}
\end{figure}

In this work, we introduce \ourname, an MRI-based foundation model developed using a large-scale self-supervised learning (SSL) approach. Our model was trained on a comprehensive MRI dataset comprising 6.9 million slices from 116,806 volumes that span virtually all anatomical regions, collected from Duke University (Fig. \ref{fig:summary}a, b). This dataset is termed ``Duke-110K'' and covers more than 9 types of MRI sequences. We chose DINOv2 as the SSL training objective due to its demonstrated effectiveness in learning rich visual representations. Notably, we adapted a nonstandard training pipeline based on our experimentation. In this strategy, the model is initialized with pre-trained weights rather than random initialization; the model is then fine-tuned for a limited number of epochs (Fig. \ref{fig:summary}c). We selected the initial weights from Segment Anything Model (SAM) \citep{kirillov2023segment} due to its impressive zero-shot segmentation capabilities across diverse imaging modalities, including MRI \citep{mazurowski2023segment}. This strategy enables faster convergence and improved representation quality. We evaluated \ourname on few-shot segmentation as well as zero-shot classification and segmentation, covering a total of 22 clinical tasks. Our results show superior performance compared to SAM and its medical variant (MedSAM \citep{ma2024segment}), underscoring the effectiveness and generalizability of the proposed approach (Fig. \ref{fig:summary}d). 

\section{Results}

\subsection{Determining the best strategy for training the foundation model}

There appears to be a general assumption that a larger model, more data, and longer training will result in a better foundation model and there is limited attention toward empirical validation of these assumptions. In the process of developing our foundation model, we conducted an ablation study to verify whether the proposed training algorithm is indeed the best one. Since our main focus is segmentation with limited training data, we based this investigation on segmentation tasks. Experiments were conducted on two in-house datasets: Private-Bone and Private-Breast. 

\begin{figure}[htbp]
    \centering
    \includegraphics[width=\linewidth]{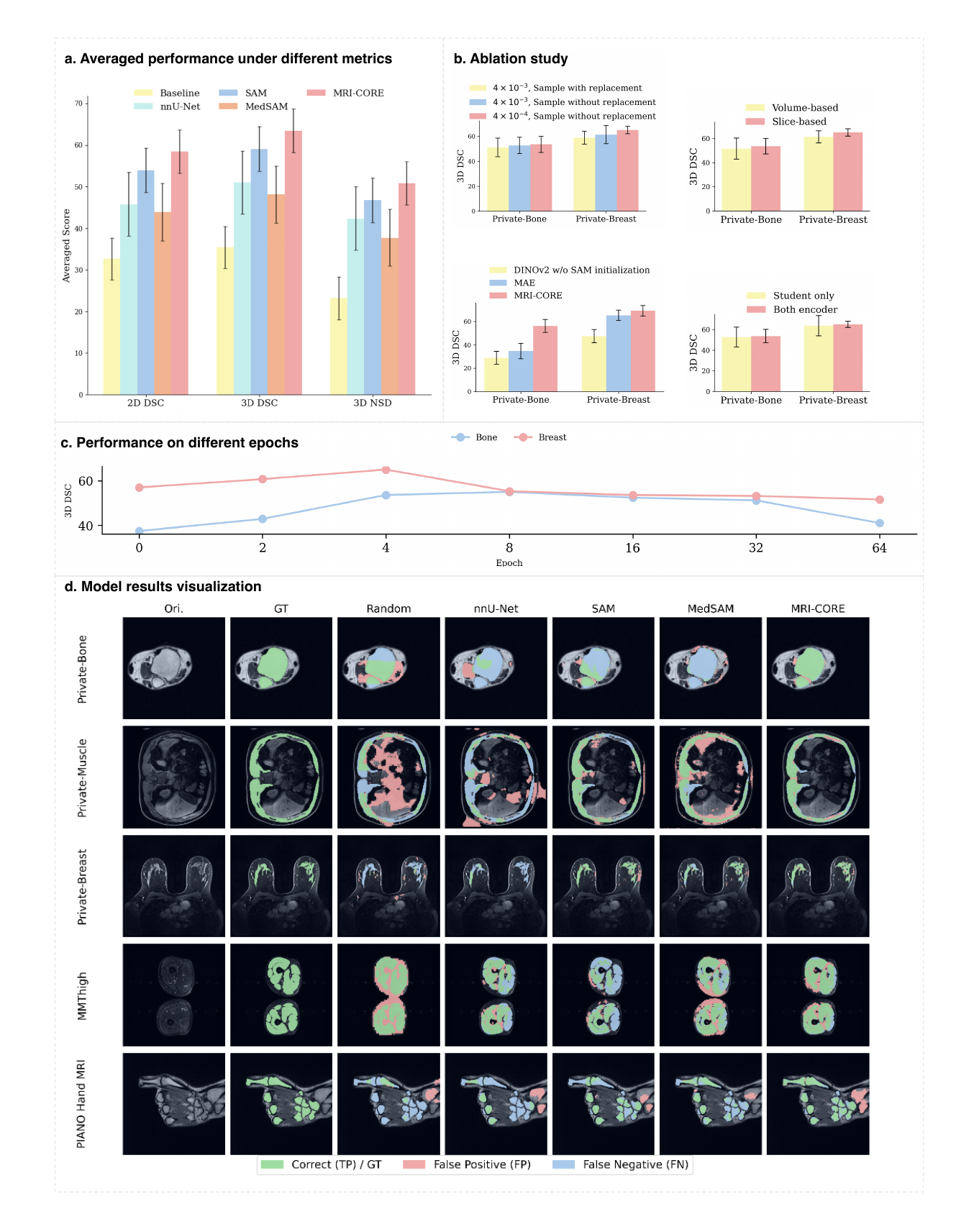}
    \caption{
    \textbf{a}. Bar plot of \ourname and all competing methods, averaged on all datasets, under different evaluation metrics. 
    \textbf{b}. Bar plot showing 3D DSC from ablation studies on core training hyperparameters, sampling strategy, normalization method, initialization scheme, and the MAE pretraining baseline. \textbf{c}. 3D DSC across different pretraining epochs on the Private-Bone and Private-Breast datasets. \textbf{d} Qualitative segmentation examples on part of evaluation datasets (see \ref{secA2} for more results), with correct predictions in green, false positives in red, and false negatives in blue.
    }
    \label{fig:result}
\end{figure}

\textbf{Segmentation Performance Across Training Epochs}. To determine the optimal training duration, we evaluated segmentation performance at different epochs. As shown in Fig. \ref{fig:result}c, the model achieves the highest 3D DSC at epoch 4. Extended training beyond this point resulted in reduced generalizability of the model showing that longer training may indeed harm model performance. Therefore, we selected the epoch 4 checkpoint for all subsequent evaluations.

\textbf{Evaluation on SAM Initialization and MAE}. We evaluated the impact of SAM initialization and the relative performance of DINOv2 versus Masked AutoEncoder (MAE) \citep{he2022masked}. Fig. \ref{fig:result}b demonstrates that DINOv2 initialized with SAM weights outperformed both DINOv2 without SAM initialization and MAE initialized with the SAM weight on both Private-Bone and Private-Breast datasets.

\textbf{Analysis of Training Hyperparameters and Settings}. Furthermore, we conducted experiments to evaluate the impact of several training factors. The details of each factor can be found in Sec. \ref{sec:abla_details}. As shown in Fig. \ref{fig:result}b, sampling without replacement outperforms DINOv2's default choice of samling with replacement since when only pre-training for a few epochs, the latter strategy implies that some images could never be selected for learning. 
A smaller learning rate of $4\times10^{-4}$ yields better performance than $4\times10^{-3}$, highlighting the effectiveness of making the learned weights similar to the initial ones. Conducting slice-wise normalization during pre-training outperforms volume-wise normalization, and initializing both encoders with SAM weights improves results compared to student-only initialization.

\subsection{Few-shot Segmentation}
% experiment setting for ablation
A key aim of our work is to improve segmentation capabilities in label-efficient settings across all MRI tasks. To this end, we consider a practical scenario where only five slices are selected for model fine-tuning and five slices for fine-tuning-related validation. We evaluated the resulting model performance using 2D Dice Similarity Coefficient (DSC), 3D DSC, and 3D Normalized Surface Dice (NSD).

% experiment setting for main
\subsubsection{Model Performance}
\label{sec:segperformance}
We conducted experiments on five common MRI segmentation tasks across 10 datasets, both non-public in-house and public: bone segmentation (Private-Bone, PIANO Hand MRI, SPIDER), muscle segmentation (Private-Muscle, MMThigh), breast tissue segmentation (Private-Breast), organ segmentation (AMOS-Liver, AMOS-Pancreas, AMOS-Spleen, QIN \& ISBI-MR, HVSMR-Chamber, HVSMR-Vessel) and tumor segmentation (LiverHccSeg-Tumor). Note that different test sets were selected for Private-Bone and Private-Breast to avoid overfitting. 
\ourname was compared to two segmentation-focused foundation models: SAM and MedSAM, and two standard approaches: nnU-Net and \textit{baseline}, which does not utilize any pre-trained weights.

As seen on Fig. \ref{fig:summary}d, \ourname achieved substantial gains in 3D DSC over existing foundation models. Compared with SAM, it increased 3D DSC by 13.9\%, 17.4\%, 2.6\%, 3.1\%, 4.7\%, 3.75\%, 6.74\%, 3.87\%, 3.12\%, 1.75\%, 4.84\% on Private-Bone, Private-Muscle, Private-Breast, PIANO Hand MRI, MMThigh, AMOS-Liver, AMOS-Pancreas, AMOS-Spleen, QIN \& ISBI-MR, HVSMR-Vessel, LiverHccSeg-Tumor datasets/tasks, respectively. Compared with MedSAM, the corresponding improvements were 27.3\%, 22\%, 13.2\%, 14.5\%, 18.1\%, 21.10\%, 10.46\%, 17.57\%, 12.08\%, 17.23\%, 3.29\% on the same benchmarks. Moreover, \ourname outperformed nnU-Net by 16.1\%, 23.7\%, 1.2\%, 4.8\%, 11.5\%, 15.11\%, 17.16\%, 33.34\%, 14.65\%, 13.32\%, 7.46\% across these datasets/tasks. Although \ourname slightly underperformed SAM on SPIDER and HVSMR-Chamber by 1.3\% and 0.7\%, respectively,  \ourname exhibited a significant margin of improvement across all evaluated metrics, including 2D DSC, 3D DSC and 3D NSD on average (Fig. \ref{fig:result}a).

Qualitatively, as shown in Fig. \ref{fig:result}c, segmentation with \ourname appears to produce more correct predictions and fewer false negatives, indicating more precise segmentation. Also, the false positives in the hand region occur exclusively on bone structures for \ourname, demonstrating that \ourname has captured relevant semantic information.

\begin{figure}[!h]
    \centering
    \includegraphics[width=0.99\linewidth]{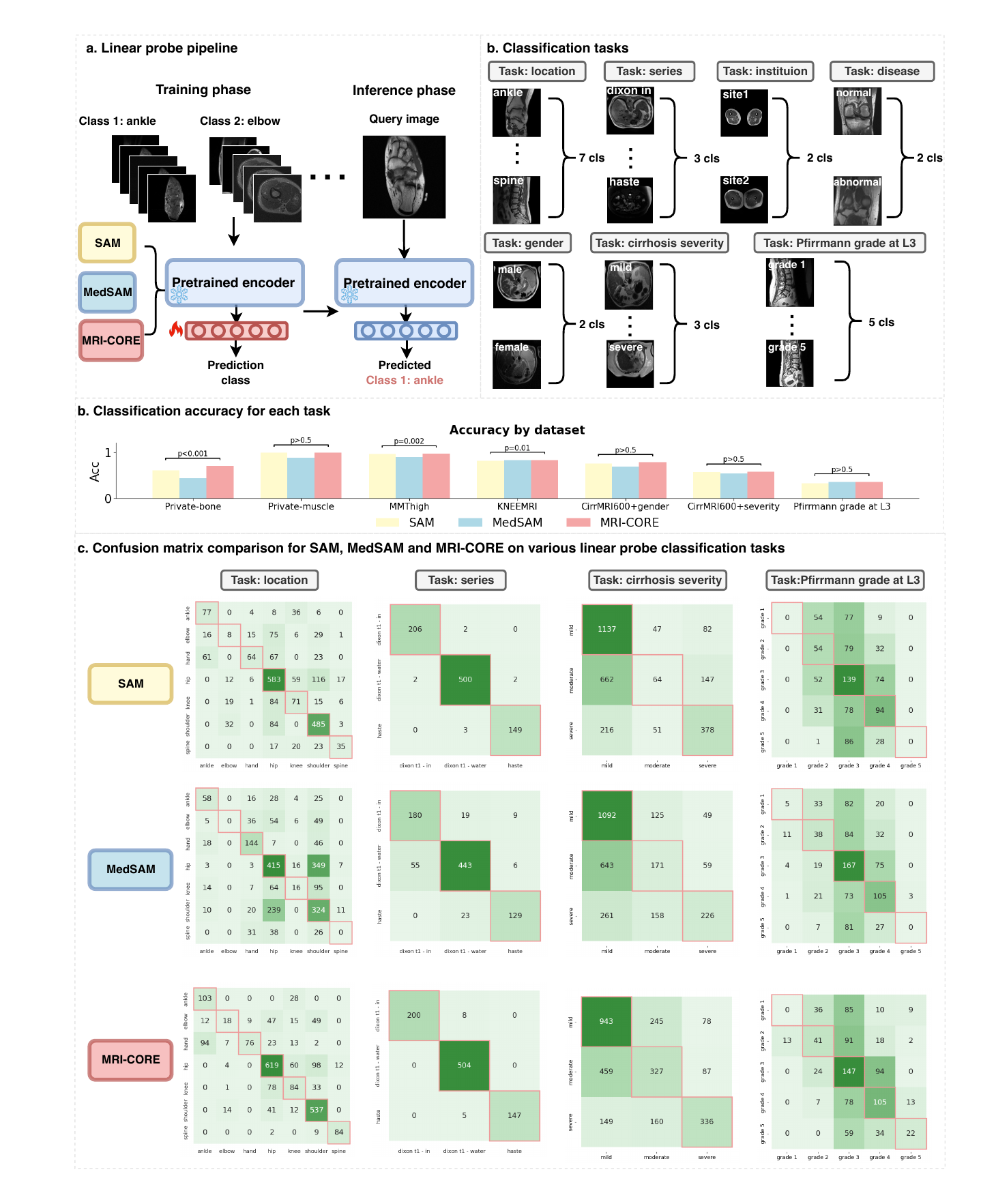}
    \caption{
    \textbf{a}. Overview of the pipeline for linear probing: the encoder is frozen and only a linear classifier is trained. \textbf{b}. Classification accuracy of SAM, MedSAM and \ourname on various linear probing tasks \textbf{c}. Confusion matrix comparison for SAM, MedSAM and \ourname on various linear probing classification tasks.
    }
    \label{fig:linear_probing}
\end{figure}

\subsubsection{Relationship Between Few-Shot Segmentation Performance and Similarity of Image Distributions}
\label{sec:FRDcorrelation}
\ourname achieves state-of-the-art (or close) performance on all surveyed MRI few-shot segmentation tasks (Fig. \ref{fig:summary}), but it is unclear \textit{why} its performance improvement over the runner-up model, SAM (\ie, \ourname before any pre-training), varied so greatly across segmentation tasks. We intuitively propose that this difference (we will refer to it as $\Delta$) is related to how similar the images in \ourname's pre-training set are to the images in the application domain, i.e., the images used for fine-tuning and testing.    
Testing this hypothesis requires comparing unpaired image distributions, and while mainstream computer vision typically uses metrics like Fr\'echet Inception Distance (FID) \citep{fid} or Kernel Inception Distance (KID) \citep{binkowski2018demystifying}, these are unsuitable for medical images as they were developed on and for natural images \citep{osuala2023data,woodland2024_fid_med,konz2024rethinking,wu2025pragmatic,frd}. We therefore use the recently proposed Fr\'echet Radiomic Distance (FRD) \citep{frd}, which compares distributions using predefined radiomic features specifically designed for medical image analysis to avoid the issues of metrics like FID, where lower values indicate higher similarity.

\begin{figure}[htbp]
    \centering
    \includegraphics[width=0.8\linewidth]{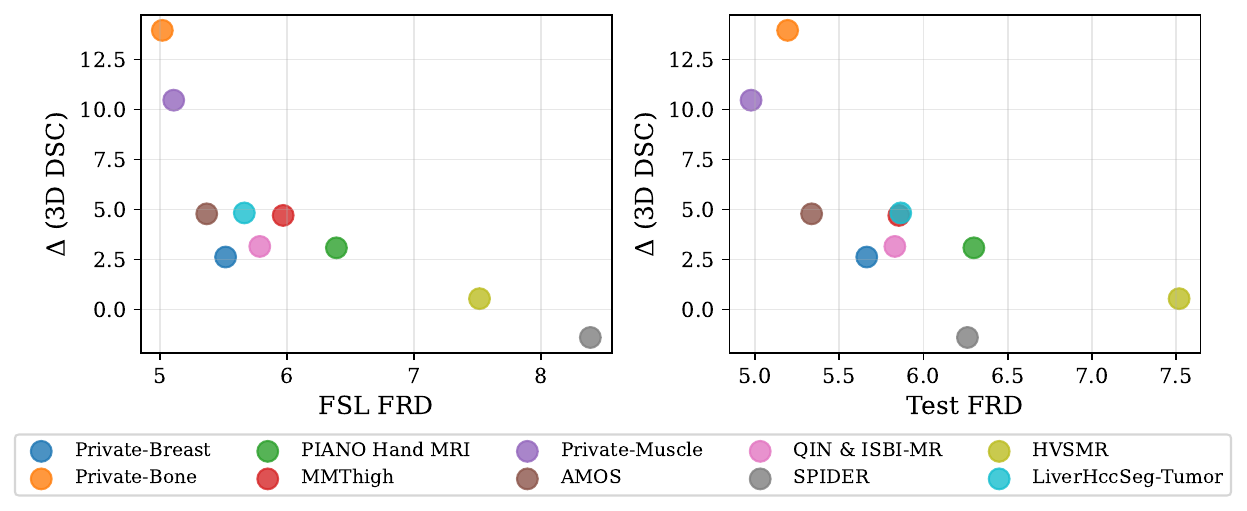}
    \caption{
    \textbf{Left:} The relationship between the change ($\Delta$) in few-show learning (FSL) segmentation performance of \ourname after pre-training from SAM, and the Fr\'echet Radiomic Distance (FRD) \citep{frd} between the pre-training set and the FSL fine-tuning set, for each segmentation dataset. \textbf{Right:} Same, but computing the FRD between the pre-training set and the test set. 
    }
    \label{fig:frdcorr}
\end{figure}

We first evaluated the relationship between $\Delta$---the few-shot performance improvement of \ourname over SAM after being fine-tuned---and the FRD between the pre-training set and the full-size datasets used to sample images for few-shot learning (FSL set). As shown in Fig. \ref{fig:frdcorr} left, each point corresponds to a particular few-show segmentation dataset (Fig. \ref{fig:summary}), where each $\Delta$ is averaged over all tasks for the given dataset if multiple exist (\ie, HVSMR and AMOS). In Fig. \ref{fig:frdcorr} right, we show similar results but with instead computing the FRD between the pre-training set and the test set itself (sampled from the same underlying distribution as the FSL set) 

The results align with our intuition: There is a strong correlation between $\Delta$ and both FRD metrics, with a Pearson linear correlation of $R_P=-0.781$ ($p=0.008$) and Spearman non-linear rank correlation $R_S=-0.842$ ($p=0.002$) between $\Delta$ and the FSL set FRD, and $R_P=-0.718$ ($p=0.02$) and $R_S=-0.721$ ($p=0.02$) between $\Delta$ and the test set FRD, which also aligns with prior results in \citet{frd} of FRD correlating with downstream task performance for various tasks. This directly establishes that the more similar a FSL fine-tuning set or test set is to the model's pre-training set, the more that the segmentation performance on the test set will improve over the original SAM.

Intuitively, the negative correlation between the FSL set and test set FRDs and $\Delta$ can be explained by the domain-specific knowledge acquired during \ourname's pre-training. When the feature distribution distance between the pre-training and downstream task is small (low FRD metrics), \ourname has already learned relevant features for similar data during pre-training, enabling more effective few-shot adaptation compared to the domain-agnostic SAM. Conversely, when FRD is high, the downstream task diverges from \ourname's pre-training distribution, diminishing the advantage of its specialized features and resulting in performance closer to SAM. The analysis in this section lays the groundwork for the analysis of a not-well-understood question of how the distribution of the training data affects the usability of foundation models in different domains.

\subsection{Classification in linear classifiers}
% experiemnt setting
In addition to evaluating \ourname's transfer ability, we also assessed its ability to capture several important properties in medical image analysis, such as location type, sequence type, acquisition site, normality, gender and disease severity/grade. To conduct this assessment, we selected Private-Bone, Private-Muscle, MMThigh, KNEEMRI, CirrMRI600+ and SPIDER to evaluate these properties. We performed logistic regression on top of the pre-extracted features from \ourname. For comparison, we extracted the outputs from MedSAM and SAM's image encoder. We evaluated all tasks using classification accuracy and reported the confusion matrix for non-binary classifications. Additional details regarding the experimental setup are provided in Section \ref{sec:lp}.

% we better on identifying image properties
As shown in Fig. \ref{fig:linear_probing}, our model outperformed the other two foundation models, SAM and MedSAM, on average in terms of classification accuracy. A detailed examination of the confusion matrix reveals that \ourname excels particularly in identifying sequence types and anatomical locations, which we attribute to the diversity and coverage provided by the Duke-110K dataset. Notably, our model achieved superior performance on classes with more subtle distinctions, such as between ankle and hand regions, surpassing SAM in these challenging cases. Furthermore, \ourname has demonstrated strong generalization capabilities to previously unseen properties, \textit{i.e.,} variations across imaging institutions, and disease severity/grade prediction, highlighting its robustness and adaptability.

% we better on identifying diseases (if true)

\subsection{Zero-shot segmentation}
\begin{figure}[htbp]
    \centering
    \includegraphics[width=\linewidth]{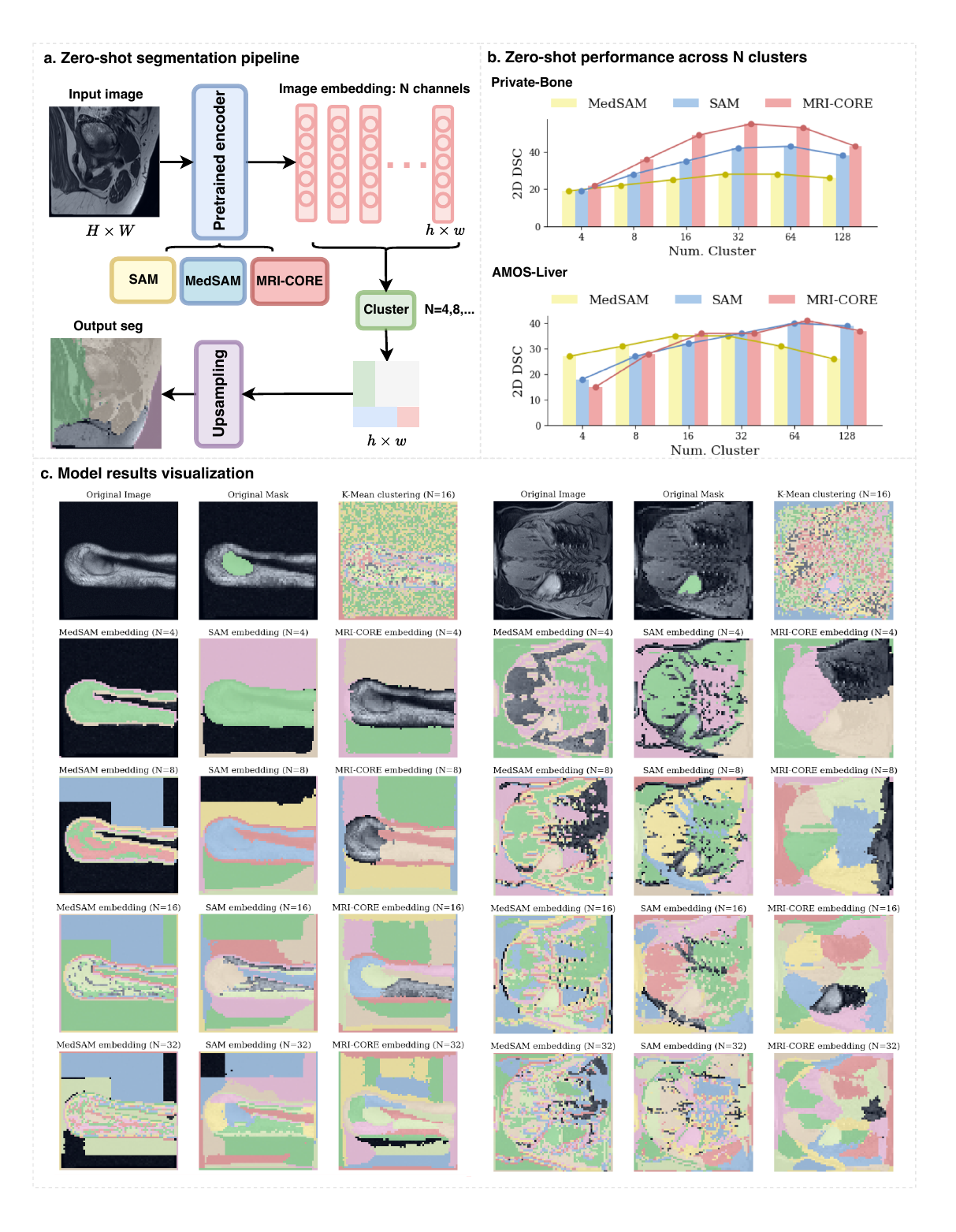}
    \caption{\textbf{a}. Overview of the zero-shot segmentation pipeline. \textbf{b}. 2D DSC of zero-shot performance on Private-Bone and AMOS-Liver across different numbers of clusters for SAM, MedSAM, and \ourname. \textbf{c}. Qualitative results on zero-shot performance for SAM, MedSAM, and \ourname under different numbers of clusters, for two examples cases (left and right sections).
    }
    \label{fig:zero_shot}
\end{figure}
% experiment setting on how to do k-means
While linear probing considers global representations of images, we are also interested in what can be extracted from the spatial feature representations of individual pixels generated by the foundation models. Toward this goal, we proposed a zero-shot segmentation pipeline by first clustering all such spatial features using K-means, and labeling each pixel according to the corresponding features' cluster. Comparison was also made to SAM and MedSAM's image encoders. A baseline version where K-means was directly applied to the original images is also included.

% analysis. MedSAM is just bad. SAM output determined by pixel intensivty. Ours determined by location, but also considers pixel intensity.
Our evaluation of the output was predominantly qualitative. 
The results in Fig. \ref{fig:zero_shot} highlight distinct differences across the methods. The baseline of performing K-means clustering simply on the raw pixels (top right images in the left and right sections of Fig. \ref{fig:zero_shot}c) yielded noisy and spatially incoherent clusters, lacking meaningful segmentation boundaries. In contrast, both SAM and MedSAM embeddings exhibited improved alignment with object structures (Fig. \ref{fig:zero_shot}); however, their clusters were still largely governed by low-level pixel similarity. Our model, on the other hand, produced clusters that were both spatially coherent and semantically meaningful, capturing the target object with greater precision and cleaner boundaries. This indicates that our learned features incorporate both spatial location and high-level semantics, making them more suitable for downstream segmentation tasks in a zero-shot manner.

% quantitative analysis. First state the setting and then some analysis on the score
Knowing location and pixel intensity allows our model's features to capture target objects that are distinct from their surroundings and have typically fixed locations, such as bones and organs. To quantify the results, we proposed to first identify images with only one object of interest, and then find the cluster that overlaps with the target the most. This allows us to approximate segmentation performance in a zero-shot setting by mimicking a one-shot label assignment scenario. As shown in Fig. \ref{fig:zero_shot}b, \ourname achieved the leading performance on Private-Bone regardless of the number of clusters. With the optimal cluster count of 32, \ourname surpassed SAM by 13. On AMOS-Liver, \ourname achieved the best performance given 64 clusters. Additionally, qualitative zero-shot segmentation results (Fig. \ref{fig:zero_shot}c) show that the output embeddings from \ourname can yield clusters that most align with the ground-truth masks compared with SAM and MedSAM, underscoring the superior ability of \ourname to capture structural semantics without task-specific training.

\section{Discussion}
% Re-summarize our algorihtm and state it's the new SOTA for few-shot segmentation and good under linear prob and zero-shot segmentation.
In this study, we have demonstrated the versatility of \ourname, a general-purpose, self-supervised foundation model for MRI. We curated Duke-110K, a pretraining dataset containing more than 6 million slice images from more than 9 sequence types and over 18 body locations. Built upon a DINOv2-based self-supervised learning approach which scales well to large datasets, we proposed a new training strategy involving carefully limited fine-tuning from pre-trained weights. \ourname has demonstrated superior performance in few-shot and zero-shot segmentation in a wide range of tasks, and has demonstrated stronger potential for classification/linear probing than SAM or MedSAM.

% start writing a detailed discussion paragraph about those points:
We found that several factors significantly influenced the performance of \ourname.  First, the choice of self-supervised learning (SSL) algorithm to pre-train \ourname played a significant role; DINOv2's student-teacher framework proved effective for our pre-training objectives, showing improvements over other SSL techniques such as MAE. Second, initializing with SAM's pre-trained weights allowed for harnessing the benefits of general pre-training of very large number of natural images and resulted in performance improvements. Third, fine-tuning for a limited number of epochs was essential to ensure that while the model adapted to medical images, it did not ``forget'' the representations learned by SAM. While this study did not include the experimental evaluation of this claim, we believe that the diversity and scale of the pre-training dataset over different body locations and sequence types were crucial, as they enabled the model to learn rich and generalizable representations, allowing it to excel in various segmentation tasks.

The ability of \ourname to excel in few-shot segmentation has important implications for medical imaging research and clinical practice. Improved few-shot learning means that high-quality segmentation models can be developed with minimal annotated data, reducing the burden on expert annotators and enabling rapid adaptation to new tasks, rare conditions, or underrepresented anatomical regions. The strong performance of \ourname in linear probing and zero-shot segmentation further suggests that the model captures generalizable and semantically meaningful features, which could facilitate the automated discovery of new imaging biomarkers, support robust classification of image properties, and enable segmentation in settings where annotated data is scarce or unavailable. Clinically, these capabilities may accelerate the deployment of AI tools in diverse healthcare environments, improve diagnostic accuracy, and broaden access to advanced image analysis, especially in resource-limited settings.

Our model, despite providing major advances in medical imaging research, has some limitations. First, the current architecture is based on ViT-Base due to computational constraints; larger models may further improve performance, but were not feasible in this study. Second, the Duke-110K dataset, while large and diverse in body regions and sequence types, is sourced from a single institution and does not include brain MRI. While this could limit generalizability, our evaluation of the model in a range of external datasets shows its robustness. Third, \ourname is currently limited to vision-only tasks and does not incorporate multi-modal data such as clinical text or other imaging modalities. Addressing these limitations in future work could further enhance the utility and impact of MRI foundation models.

The broader research impact of \ourname lies in its potential to serve as a foundation for future advances in medical image analysis. By making the model and codebase publicly available under a permissive license, we aim to lower the barrier for others to develop, adapt, and deploy high-performing segmentation and classification models for MRI. This democratization of foundation model technology could catalyze new research directions, foster reproducibility, and accelerate the translation of AI innovations into clinical workflows. We envision \ourname as a new standard for MRI-based vision tasks, providing a robust starting point for both academic and translational research.

\section{Methods}\label{sec11}

\subsection{Curation of Duke-110K}
We curated a diverse MRI dataset through a structured, multi-stage pipeline (Fig. \ref{fig:summary}a). The data was collected under an institutional review board protocol at Duke University Medical Center. First, we identified all MRI-related radiology reports within the clinical data portal between 2016 and 2020. We then extracted the corresponding ``Report Type'' for each report (e.g., MRI left shoulder with and without contrast) to determine all available MRI protocols. We example protocols across body locations and exclude those that generally contain facial anatomy—such as brain, face, and neck MRIs—and retained 107 unique MRI protocol categories. For each protocol, we randomly sampled 150 exams, ensuring one exam per unique patient to avoid redundancy and maintain statistical independence. We included all available exams if a protocol had fewer than 150 exams captured within the 5-year range. The selected exams were then retrieved using their accession numbers and downloaded from our image repository. They were then de-identified to remove patient information. 
We then discarded any sets that failed to form a continuous volume—such as those missing slices or reference slices. The remaining DICOMs represented 116,806 distinct MRI volumes, covering a wide range of body locations (e.g., abdomen, hip, thigh, breast), sequence types (e.g., T1, T2, STIR), and view orientations (axial, coronal, sagittal), as shown in Fig. \ref{fig:summary}b.
Finally, we converted all DICOMs into 2D PNG slices with min-max normalization. 

\subsection{Large-scale visual pre-training}
\subsubsection{Algorithm details}
In this work, we used DINOv2 \citep{oquab2023dinov2}, a state-of-the-art self-supervised learning method based on student-teacher knowledge distillation, for pre-training the network. DINOv2 combines two previous methods, DINO \citep{caron2021emerging} and iBOT \citep{zhou2021ibot}. DINO is a self-distillation loss where two different random augmentations of an input image are fed to the student and teacher networks, two networks with identical structures, respectively. The student network is trained to match the output from the teacher network, and the teacher network is a slowly updated exponential moving average of past student networks. iBOT employs a reconstruction-based loss where random input patches are masked for the student but remain visible to the teacher. The student encoder predicts features for the masked tokens, which are then matched against the corresponding visible features from the teacher encoder.

The core difference between our approach and the original DINOv2 pipeline lies in the initialization strategy. Instead of training the model from scratch, we initialized it with pre-trained weights. Specifically, we adapted SAM’s image encoder, motivated by SAM's strong zero-shot segmentation performance on the natural image domain. To integrate SAM into our framework, we replaced both the student and teacher encoders in DINOv2 with SAM’s pre-trained image encoder. Under this setting, we observed that extended training can negatively affect the performance. While the default DINOv2 schedule involves 1,250,000 training iterations, we achieved the best results with only 4 epochs (approximately 50,000 iterations).

\ourname utilized the ViT-B architecture and adopted SAM's ViT-B backbone. We used base learning rate of $4\times 10^{-4}$. We select epoch-based sampling strategy (all sampled are fed to network before next epoch run). Batch size of 128 per GPU and trained on 4 NVIDIA A6000 GPUs. Training 1 epoch took approximately 3 hours.

\subsubsection{Ablation study design} \label{sec:abla_details}
To validate the design choice of our algorithm, we performed ablation studies on several important variables that were identified to have an effect on the performance. All results were evaluated on Private-Bone and Private-Breast datasets. Note that the evaluation test set was different from the real test set to avoid overfitting.
The most important variable was the number of training epochs. We report the performance on 2, 4, 8, 16, 32, and 64 epochs to observe the changes.
Next, we discovered that under this setting, the current sampling strategy (i.e., sampling without replacement to ensure no images are resampled within an epoch) is better than sampling with replacement, and a smaller learning rate is beneficial.
When initializing the network weights, we compare applying SAM-based initialization to both student and teacher encoders versus only applying it to the student encoder. We omitted the ``teacher-only'' baseline since under the EMA update ($w_{new} \approx 0.999^{iter\_num} w_{old}$), as the teacher would rapidly inherit the student's weights over prolonged training. Furthermore, we compared our training strategy with two variations: (1) plain DINOv2 without pre-training weights and (2) MAE. Each method was trained up to 100 epochs. We evaluated at an interval of 20 epochs and reported the best one. Lastly, we investigated whether using different data normalization strategies, \textit{i.e.,} slice-based normalization or volume-based normalization, during pre-training would affect the performance.  

\subsection{Evaluation setting}

\subsubsection{Few-shot segmentation} \label{sec:fss}
We adopted the optimal fine-tuning strategy proposed by \citep{gu2024build}, which involves inserting Adapter layers \citep{houlsby2019parameter} into the first two and last two transformer blocks of the image encoder. Each Adapter layer consists of two linear functions and a nonlinearity. All input images were resized to $1024 \times 1024$ to comply with SAM's input requirements, and min-max normalization was applied per slice. The base learning rate was $10^{-4}$. Data augmentation strategies included color jittering and random resizing and cropping. Models were trained for at least 1000 epochs and used early stopping if the validation score did not improve for 200 epochs. This extensive training ensures the model can sufficiently adapt to the limited examples available in the few-shot setting.

\subsubsection{Linear probing protocol} \label{sec:lp}
Following the DINOv2 protocol, we performed linear probing by first extracting patch-level features from the frozen image encoder. These features were aggregated via 2D average pooling across spatial dimensions to obtain a single feature vector per image. A linear classifier was then trained using the Adam optimizer with L2 regularization for 100 epochs. We used a learning rate of $10^{-4}$ and optimize with cross-entropy loss. No data augmentations were applied. All images were resized to $1024 \times 1024$ and min-max normalized per slice. We reported the best validation performance across training epochs, consistent with the DINOv2 evaluation protocol. For each task, data was split by patient into train:val sets at a 60\%:40\% ratio to best maintain label balance, and the best validation performance was reported, following \citep{oquab2023dinov2}.

\subsubsection{Zero-shot segmentation} \label{sec:zss}
For zero-shot segmentation, we clustered the extracted image features using k-means from scikit-learn. The k-means algorithm was initialized with init='auto' and a fixed random seed for reproducibility. The number of clusters (k) was treated as a hyperparameter. We experimented with k = 4, 8, 16, and 32 for visual inspection, and additionally included k = 64 and 128 for quantitative evaluation. Each cluster was mapped to a semantic region based on majority voting over the pixel-wise or region-wise labels in the ground truth.

\subsection{Datasets used for evaluation} \label{sec:dsdata}
In this section, we outline the data preprocessing, number of samples per class, train–validation–test folds, and other details per dataset (which may also span multiple tasks). In this study, we have selected 3 internal datasets and 9 public datasets for evaluation. All but the bone and breast datasets were split into the training, validation, and test sets. The bone and breast datasets were split into four sets, including an additional development test set for the ablation study.
%\todo{add which datasets for seg, which for cls.}

\textbf{Private-Bone dataset} includes 8,534 T1-weighted MRI, 2D slices from 106 patients and spanning 16 anatomical regions. The dataset stems from the SegmentAnyBone project \citep{gu2025segmentanybone}. Each slice contains binary bone segmentation that has been rigorously reviewed by a senior musculoskeletal radiologist and an orthopedic surgery fellow. Bone segmentation is particularly challenging in this setting due to the wide variety in both intensities and shapes of bones across different body regions. The dataset was split into 6,134:1,369:1,031 slices for training, validation, and testing by patient. For linear probing, we performed body-location classification by merging labels as follows: lumbar spine and thoracic spine into ``spine''; hip, pelvis, and thigh into ``hip''; shoulder and humerus into ``shoulder''; elbow and forearm into ``elbow''; hand and wrist into ``hand''; knee and lower leg into ``knee''; and ankle and foot into ``ankle''.

\textbf{Private-Breast dataset} comprises 973 slices acquired from 58 patients, featuring individuals with diverse breast densities and morphologies, including cases with unilateral presentations. To maximize the range of segmentable structures, segmentation was performed on the first post-contrast volume. The entire dataset underwent rigorous review and confirmation by experienced radiologists. For this work, segmentation targeted the dense breast tissue within this dataset, utilizing it as a mask. The accurate segmentation of dense breast tissue is inherently challenging. This difficulty stems from its often heterogeneous texture, indistinct boundaries, and variable appearance, which can complicate precise delineation. The dataset was split into 559:169:245 slices for training, validation, and testing, by patient.

\textbf{Private-Muscle dataset} comprises 7,407 MRI 2D slices spanning over 11 anatomical regions and multiple imaging modalities. Each slice includes a binary muscle segmentation mask that has been reviewed and approved by a senior musculoskeletal radiologist. Accurate muscle segmentation is particularly difficult due to variations in tissue appearance and shapes across different body regions. The dataset was split into 5,189:1,110:1,108 slices for training, validation, and testing, by patient. For linear probing, we selected three abdominal sequence types: Dixon T1-in, Dixon T1-water and Haste.

\textbf{PIANO Hand MRI dataset} is collected from \citep{li2021piano} and comprises 4,089 2D slices extracted from 50 T1-weighted hand MRI scans. Accurate segmentation of hand bone is essential for realistic hand modeling and biomechanics analysis, with potential downstream applications in virtual and augmented reality. Each slice includes bone segmentation annotations, enabling evaluation on a bone segmentation task. The complexity arises from the fine-scale bone geometry and marked variability in hand posture and anatomical presentation. The dataset was split into 2,137:908:1,044 slices for training, validation, and testing, by patient.

\textbf{Multimodal Multiethnic Thigh Muscle MRI Analysis (MMThigh) dataset} comprises 5,358 2D axial leg slices extracted from 93 MRI volumes, collected across 2 institutions (Helsinki: 27 volumes and HuashanMyo: 66 volumes) and spanning 5 MRI modalities, including STIR, T1, IDEAL Fat, T2, and IDEAL Water \citep{thighmuscle}. This dataset includes manual segmentation of thigh muscles, and we perform binary muscle segmentation. The heterogeneous modalities and multi-site acquisitions introduce substantial variability in contrast and tissue appearance. The dataset was split into 3,276:576:1,506 slices for training, validation, and testing, by patient. We conducted institutional classification tasks for linear probing on MMThigh.

\textbf{AMOS MRI dataset} is collected from \citep{ji2022amos}, comprising 5,615 2D slices extracted from 60 MRI volumes. Each volume is annotated at the voxel level for 15 abdominal organs. Accurate segmentation of organs such as liver, pancreas and spleen is critical for applications in disease diagnosis, surgical planning and radiotherapy, yet remains challenging due to fine-scale organ morphology and low-contrast boundaries between adjacent structures. The dataset was split into 3,831:596:1,188 slices for training, validation, and testing, by volume. We performed popular organ segmentation tasks, including liver, pancreas, and spleen, on AMOS MRI.

\textbf{QIN \& ISBI-MR Prostate dataset} is a combination of two publicly available prostate datasets, QIN-Prostate \citep{fedorov2016data} and ISBI-MR-Prostate \citep{bloch2015nci}. The dataset comprises 1,662 2D axial slices from 70 volumes (QIN: 30 volumes, ISBI-MR: 40 volumes). Each slice includes expert delineations of the central gland. Central gland segmentation is particularly challenging due to its low contrast against peripheral zone, irregular boundary morphology, and inter-patient variability in prostate size and pathology. The combined dataset was split into 1168:163:331 slices for training, validation, and testing, by patient.

\textbf{SPIDER MRI Lumbar dataset} comprises 13,838 sagittal slices extracted from 440 volumes spanning over both T1 and T2 sequences \citep{van2024lumbar}. Each slice includes annotations of lumbar vertebra. Vertebra segmentation in this dataset is particularly challenging due to variability in tissue contrast between T1 and T2 sequences and inter-subject anatomical variability. The dataset was split into 10120:1166:2552 slices for training, validation, and testing, by patient. In addition to segmentation, we conducted L3-level Pfirrman grade classification on SPIDER dataset, which includes 5 classes from 1 to 5. Pfirrmann label can reflect the disc degeneration severity, which is an important biomarker for back pain assessment, treatment planning, and longitudinal studies of spinal health.

\textbf{HVSMR dataset} comprises 8,329 2D slices from 60 cardiovascular magnetic resonance volumes. Each slice includes expert annotations of the four cardiac chambers (left ventricle, right ventricle, left atrium, right atrium) and  four vessels (aorta, pulmonary artery, superior vena cava and inferior vena cava), and we perform segmentation tasks on chamber and vessels by merging the respective labels. Cardiac segmentation is particularly difficult due to the low contrast between myocardium and blood pool. Accurate delineation of cardiac structures is clinically significant and essential for diagnosing heart failure and evaluating valvular disease. The dataset was split into 5,925:778:1626 slices for training, validation, and testing, by patient.

\textbf{LiverHccSeg dataset} includes 1,378 2D slices from 17 MRI volumes \citep{gross2023liverhccseg}. Each volume includes annotations of hepatocellular carcinoma (HCC) tumors at voxel level. Accurate segmetnation of HCC tumors is essential for the diagnosis, treatment, and monitoring of HCC patients. However, this task is difficult due to heterogeneous contract of HCC nodules, and variable lesion shape and size. The dataset was split into 783:280:387 slices for training, validation and testing, by patient.

\textbf{CirrMRI600+ dataset} comprises 6,766 2D axial MRI slices extracted from 318 volumes \citep{jha2024cirrmri600}. Each slice is labeled for both patient gender (male, female) and radiologically assessed liver cirrhosis severity (mild, moderate, severe). We performed linear probe classification tasks on gender and disease severity.

\textbf{KNEEMRI dataset} comprises 4,554 2D slices extracted from 1,217 MRI volumes \citep{kneemri2024kneeMRI}. Each slice is annotated as either normal or abnormal, supporting pathology detection. We conducted linear probe classification task on normality.

\section{Data availability}
Public datasets analyzed in this work can be accessed in their respective data websites: PIANO Hand MRI dataset (\url{https://github.com/reyuwei/PIANO_mri_data}), MMThigh (\url{https://github.com/Hirriririir/Multimodal-Multiethnic-Thigh-Muscle-MRI-analysis}), AMOS-MRI (\url{https://zenodo.org/records/7155725}), QIN-Prostate (\url{https://www.cancerimagingarchive.net/collection/qin-prostate/}), ISBI-MR-Prostate (\url{https://www.cancerimagingarchive.net/analysis-result/isbi-mr-prostate-2013/}), SPIDER (\url{https://zenodo.org/records/8009680}), HVSMR (\url{https://figshare.com/collections/HVSMR-2_0_A_3D_cardiovascular_MR_dataset_for_whole-heart_segmentation_in_congenital_heart_disease/7074755/2}), LiverHccSeg (\url{https://zenodo.org/records/7957516}), CirrMRI600+ (\url{https://osf.io/cuk24/}), and KNEEMRI (\url{https://universe.roboflow.com/kneemri-rqdae/kneemri-apmcr}). Restrictions apply to the availability of the developmental and validation datasets. De-identified data may be available for research purposes from the corresponding authors upon reasonable request.

\section{Code availability}
The code to reproduce the results and model weights can be accessed at \href{https://github.com/mazurowski-lab/mri_foundation}{https://github.com/mazurowski-lab/mri\_foundation}.

%\backmatte

%\bmhead{Supplementary information}

%\bmhead{Acknowledgment}

%\section*{Declarations}


\newpage
\bibliography{sn-bibliography}% common bib file

\newpage
\begin{appendices}

\section{Few-shot segmentation's quantitative performance}\label{secA1}

Tables \ref{tab:private_bone} - \ref{tab:tumor} show the performance of \ourname and all competing methods on the 10 evaluated datasets respectively.

\begin{table}[h]
    \centering
    \begin{tabular}{c|ccc}
       Method & 2D DSC & 3D DSC & 3D NSD \\
       \hline
       Baseline & 24.86 (1.99) & 27.17 (1.89) & 19.59 (4.75) \\
       nnU-Net  & 40.17 (12.9) & 40.31 (13.6) & 43.22 (11.9) \\
       SAM      & 39.47 (6.10) & 42.28 (6.42) & 40.97 (6.44) \\
       MedSAM   & 26.95 (5.24) & 28.90 (6.06) & 33.53 (7.83) \\
       \ourname     & \textbf{54.54 (5.38)} & \textbf{56.24 (5.58)} & \textbf{54.44 (5.62)} \\
    \end{tabular}
    \caption{Results on the Private-Bone dataset (mean and std).}
    \label{tab:private_bone}
\end{table}

\begin{table}[h]
    \centering
    \begin{tabular}{c|ccc}
       Method & 2D DSC & 3D DSC & 3D NSD \\
       \hline
       Baseline & 46.00 (1.67) & 49.73 (3.46) & 39.56 (5.24) \\
       nnU-Net  & 40.68 (8.93) & 47.58 (10.8) & 44.80 (5.58) \\
       SAM      & 46.90 (5.25) & 52.45 (5.53) & 47.58 (2.44) \\
       MedSAM   & 45.28 (5.84) & 47.92 (6.81) & 43.72 (1.92) \\
       \ourname     & \textbf{58.42 (6.92)} & \textbf{62.92 (6.14)} & \textbf{54.79 (3.50)} \\
    \end{tabular}
    \caption{Results on the Private-Muscle dataset (mean and std).}
    \label{tab:private_muscle}
\end{table}

\begin{table}[h]
    \centering
    \begin{tabular}{c|ccc}
       Method & 2D DSC & 3D DSC & 3D NSD \\
       \hline
       Baseline & 34.50 (5.76) & 35.54 (5.73) & 43.05 (7.77) \\
       nnU-Net  & 61.70 (3.48) & 64.97 (4.79) & 74.47 (5.00) \\
       SAM      & 65.06 (2.83) & 66.62 (4.01) & 78.07 (4.63) \\
       MedSAM   & 53.48 (7.33) & 56.01 (6.33) & 69.37 (5.04) \\
       \ourname     & \textbf{66.59 (4.10)} & \textbf{69.25 (4.44)} & \textbf{79.95 (3.51)} \\
    \end{tabular}
    \caption{Results on the Private-Breast dataset (mean and std).}
    \label{tab:private_breast}
\end{table}

\begin{table}[h]
    \centering
    \begin{tabular}{c|ccc}
       Method & 2D DSC & 3D DSC & 3D NSD \\
       \hline
       Baseline & 32.99 (4.27) & 35.61 (5.29) & 30.45 (4.22) \\
       nnU-Net  & 64.58 (7.75) & 67.94 (10.6) & 67.76 (10.0) \\
       SAM      & 64.76 (5.20) & 69.62 (6.01) & 66.69 (6.12) \\
       MedSAM   & 53.66 (5.74) & 58.21 (6.51) & 54.65 (5.99) \\
       \ourname     & \textbf{68.88 (3.76)} & \textbf{72.71 (4.33)} & \textbf{69.39 (5.62)} \\
    \end{tabular}
    \caption{Results on the PIANO Hand MRI dataset (mean and std).}
    \label{tab:public_hand}
\end{table}

\begin{table}[h]
    \centering
    \begin{tabular}{c|ccc}
       Method & 2D DSC & 3D DSC & 3D NSD \\
       \hline
       Baseline & 56.85 (2.53) & 56.61 (2.62) & 38.25 (4.16) \\
       nnU-Net  & 64.86 (5.74) & 67.05 (5.88) & 63.30 (5.62) \\
       SAM      & 69.25 (2.29) & 71.97 (2.32) & 65.87 (4.31) \\
       MedSAM   & 57.29 (6.39) & 58.61 (6.73) & 54.24 (4.30) \\
       \ourname     & \textbf{74.31 (1.82)} & \textbf{76.68 (2.02)} & \textbf{70.03 (3.34)} \\
    \end{tabular}
    \caption{Results on the MMThigh dataset (mean and std).}
    \label{tab:public_leg}
\end{table}

\begin{table}[h]
    \centering
    \begin{tabular}{c|ccc}
       Method & 2D DSC & 3D DSC & 3D NSD \\
       \hline
       Baseline & 48.69 (4.40) & 51.71 (6.04) & 14.88 (3.31) \\
       nnU-Net  & 62.45 (4.12) & 70.16 (3.85) & 37.09 (4.25) \\
       SAM      & 74.88 (2.47) & 81.51 (1.77) & 45.54 (3.84) \\
       MedSAM   & 61.29 (5.30) & 64.17 (6.93) & 29.12 (4.89) \\
       \ourname     & \textbf{78.48 (1.45)} & \textbf{85.27 (1.15)} & \textbf{52.34 (2.94)} \\
    \end{tabular}
    \caption{Results on the AMOS-Liver dataset (mean and std).}
    \label{tab:amos_liver}
\end{table}

\begin{table}[h]
    \centering
    \begin{tabular}{c|ccc}
       Method & 2D DSC & 3D DSC & 3D NSD \\
       \hline
       Baseline & 3.56 (3.67) & 3.76 (4.20) & 2.71 (2.86) \\
       nnU-Net  & 14.72 (6.42) & 16.43 (6.78) & 13.15 (4.40) \\
       SAM      & 24.19 (5.25) & 26.85 (4.93) & 19.24 (3.58) \\
       MedSAM   & 21.26 (5.60) & 23.13 (6.33) & 17.15 (3.56) \\
       \ourname     & \textbf{30.41 (5.48)} & \textbf{33.59 (5.12)} & \textbf{22.99 (3.53)} \\
    \end{tabular}
    \caption{Results on the AMOS-Pancreas dataset (mean and std).}
    \label{tab:amos_pancreas}
\end{table}

\begin{table}[h]
    \centering
    \begin{tabular}{c|ccc}
       Method & 2D DSC & 3D DSC & 3D NSD \\
       \hline
       Baseline & 19.97 (7.11) & 22.62 (8.02) & 8.21 (3.41) \\
       nnU-Net  & 27.44 (6.84) & 36.12 (9.30) & 23.80 (5.36) \\
       SAM      & 61.12 (7.08) & 65.59 (6.69) & 39.26 (9.91) \\
       MedSAM   & 48.30 (11.89) & 51.89 (9.44) & 30.65 (5.73)  \\
       \ourname     & \textbf{64.16 (8.30)} & \textbf{69.46 (6.76)} & \textbf{45.98 (7.39)} \\
    \end{tabular}
    \caption{Results on the AMOS-Spleen dataset (mean and std).}
    \label{tab:amos_spleen}
\end{table}

\begin{table}[h]
    \centering
    \begin{tabular}{c|ccc}
       Method & 2D DSC & 3D DSC & 3D NSD \\
       \hline
       Baseline & 36.41 (8.20) & 36.53 (7.31) & 23.59 (4.36) \\
       nnU-Net  & 55.36 (10.55) & 57.69 (9.84) & 43.45 (9.55) \\
       SAM      & 67.85 (3.60) & 69.22 (3.83) & 51.18 (5.31) \\
       MedSAM   & 58.80 (6.59) & 60.26 (5.34) & 44.92 (4.39) \\
       \ourname     & \textbf{71.53 (4.05)} & \textbf{72.34 (4.19)} & \textbf{56.65 (6.22)} \\
    \end{tabular}
    \caption{Results on the QIN \& ISBI-MR dataset (mean and std).}
    \label{tab:prostate}
\end{table}

\begin{table}[h]
    \centering
    \begin{tabular}{c|ccc}
       Method & 2D DSC & 3D DSC & 3D NSD \\
       \hline
       Baseline & 26.61 (5.06) & 36.34 (6.43) & 32.51 (5.04) \\
       nnU-Net  & 49.60 (5.05) & 68.67 (7.99) & 63.48 (8.00) \\
       SAM      & 52.52 (3.61) & \textbf{72.48 (4.82)} & 69.71 (6.41) \\
       MedSAM   & 40.95 (4.67) & 56.69 (6.65) & 54.09 (4.76) \\
       \ourname     & \textbf{52.33 (5.06)} & 71.19 (7.86) & \textbf{69.93 (7.83)} \\
    \end{tabular}
    \caption{Results on the SPIDER dataset (mean and std).}
    \label{tab:spider}
\end{table}

\begin{table}[h]
    \centering
    \begin{tabular}{c|ccc}
       Method & 2D DSC & 3D DSC & 3D NSD \\
       \hline
       Baseline & 62.34 (3.45) & 66.55 (2.12) & 29.99 (6.04) \\
       nnU-Net  & 65.72 (4.99) & 70.18 (4.24) & 40.00 (2.96) \\
       SAM      & \textbf{75.44 (4.87)} & \textbf{78.88 (4.51)} & \textbf{44.27 (6.85)} \\
       MedSAM   & 57.66 (11.02) & 63.22 (9.15) & 31.68 (6.29) \\
       \ourname     & 74.14 (3.65) & 78.22 (2.99) & 42.34 (3.39) \\
    \end{tabular}
    \caption{Results on the HVSMR-Chamber dataset (mean and std).}
    \label{tab:chamber}
\end{table}

\begin{table}[h]
    \centering
    \begin{tabular}{c|ccc}
       Method & 2D DSC & 3D DSC & 3D NSD \\
       \hline
       Baseline & 28.36 (10.31) & 35.07 (12.40) & 17.58 (8.14) \\
       nnU-Net  & 36.87 (11.03) & 45.57 (12.87) & 30.59 (8.57) \\
       SAM      & 46.38 (8.45) & 57.14 (10.16) & 32.85 (6.86) \\
       MedSAM   & 32.05 (11.26) & 41.67 (11.64) & 21.47 (6.30) \\
       \ourname     & \textbf{47.81 (5.63)} & \textbf{58.89 (6.53)} & \textbf{33.23 (4.79)} \\
    \end{tabular}
    \caption{Results on the HVSMR-Vessel dataset (mean and std).}
    \label{tab:vessel}
\end{table}

\begin{table}[h]
    \centering
    \begin{tabular}{c|ccc}
       Method & 2D DSC & 3D DSC & 3D NSD \\
       \hline
       Baseline & 2.86 (6.35) & 2.92 (6.48) & 1.08 (2.06) \\
       nnU-Net  & 10.80 (10.73) & 10.53 (12.43) & 4.99 (4.12) \\
       SAM      & 13.20 (10.29) & 13.16 (9.82) & 6.15 (4.27) \\
       MedSAM   & 13.79 (12.41) & 14.71 (12.99) & 5.59 (5.83) \\
       \ourname     & \textbf{17.92 (12.87)} & \textbf{17.99 (12.88)} & \textbf{8.39 (6.78)} \\
    \end{tabular}
    \caption{Results on the LiverHccSeg-Tumor dataset (mean and std).}
    \label{tab:tumor}
\end{table}

\clearpage
% \section{Few-shot segmentation's qualitative performance}\label{secA2}

% \begin{figure}[htbp]
%     \centering
%     \includegraphics[width=0.75\linewidth]{appendix_vis.pdf}
%     \caption{Qualitative results for few-shot segmentation on each dataset.
%     }
%     \label{fig:few_shot_vis}
% \end{figure}

\section{Few-shot segmentation's qualitative performance}
\label{secA2}

\subsection{Private-Bone}
\begin{figure}[htbp]
    \centering
    \includegraphics[width=0.9\linewidth]{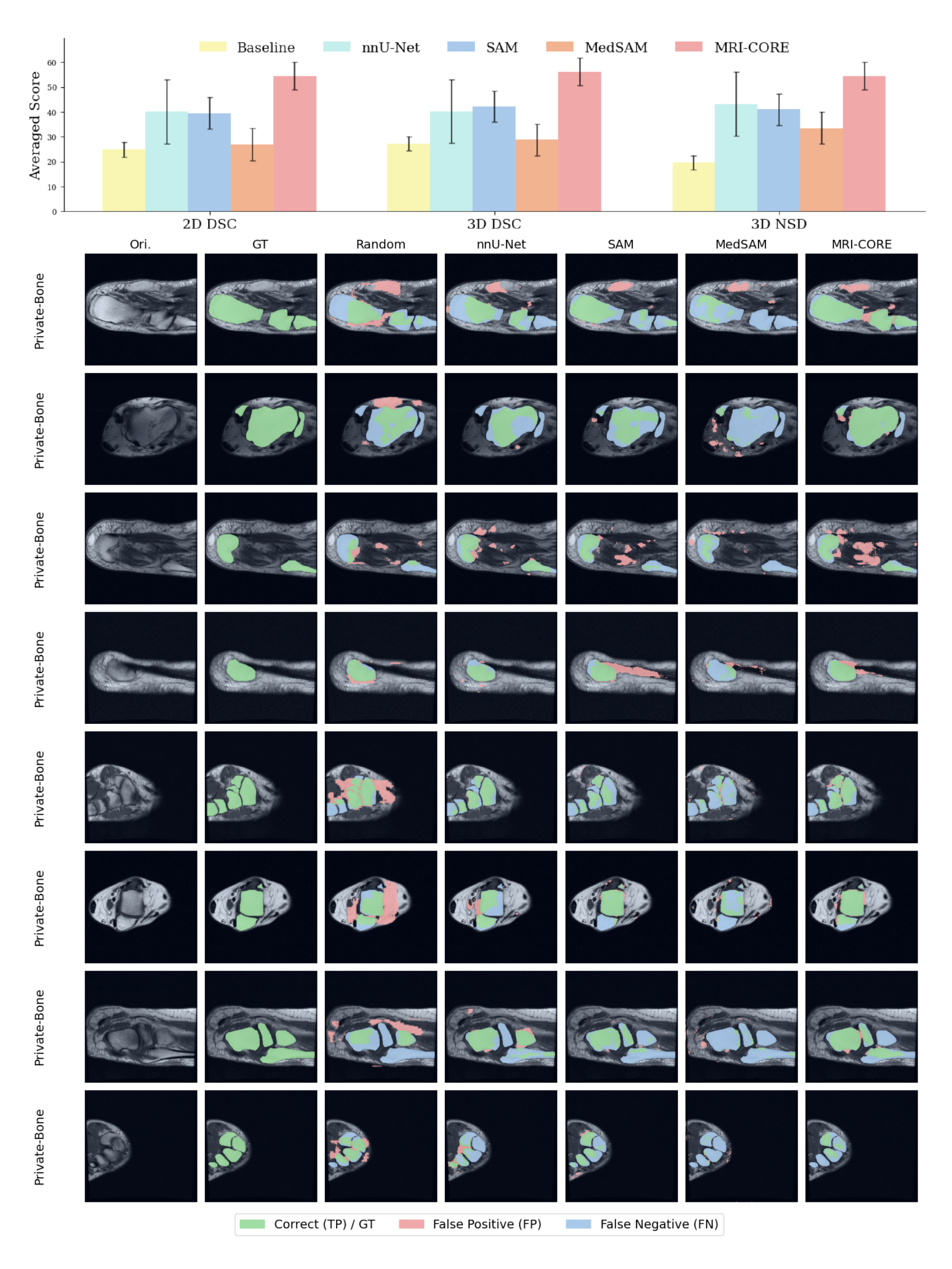}
    \caption{Qualitative results for few-shot segmentation on Private-Bone dataset.
    }
    \label{fig:app_bone}
\end{figure}
\clearpage

\subsection{Private-Muscle}
\begin{figure}[htbp]
    \centering
    \includegraphics[width=0.9\linewidth]{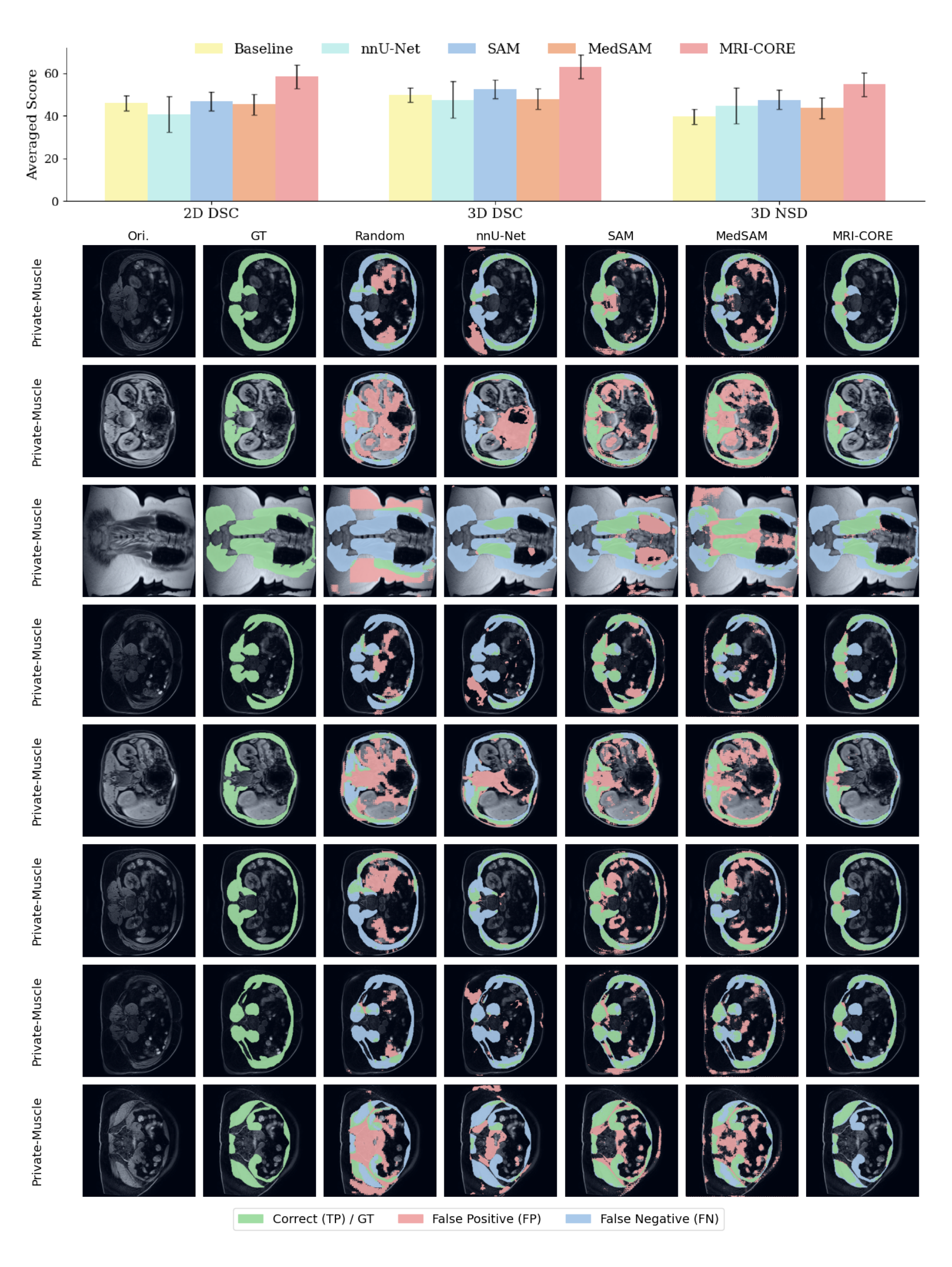}
    \caption{Qualitative results for few-shot segmentation on Private-Muscle dataset.
    }
    \label{fig:app_muscle}
\end{figure}
\clearpage

\subsection{Private-Breast}
\begin{figure}[htbp]
    \centering
    \includegraphics[width=0.9\linewidth]{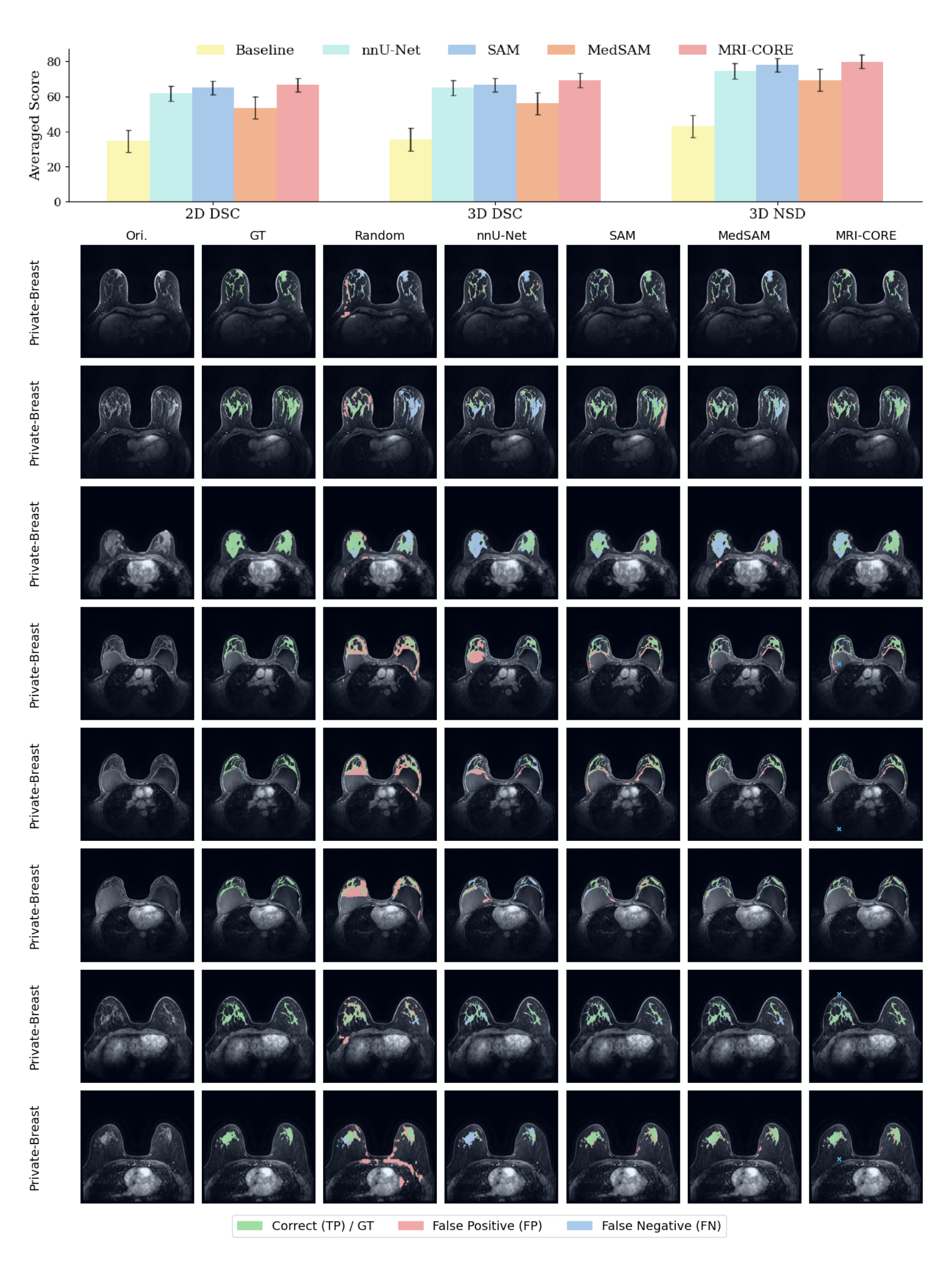}
    \caption{Qualitative results for few-shot segmentation on Private-Breast dataset.
    }
    \label{fig:app_breast}
\end{figure}
\clearpage

\subsection{PIANO Hand MRI}
\begin{figure}[htbp]
    \centering
    \includegraphics[width=0.9\linewidth]{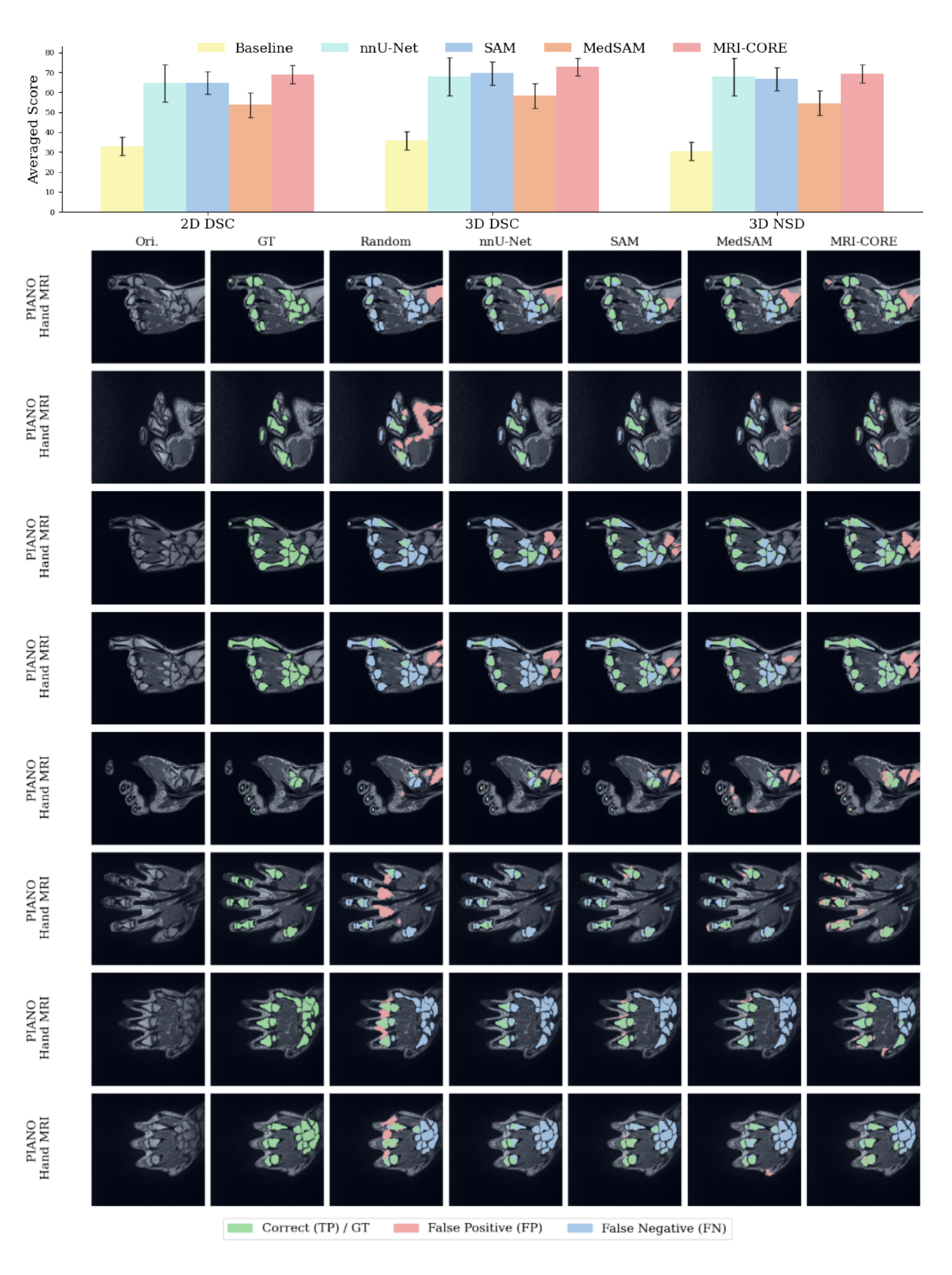}
    \caption{Qualitative results for few-shot segmentation on PIANO Hand MRI dataset.
    }
    \label{fig:app_hand}
\end{figure}
\clearpage

\subsection{MMThigh}
\begin{figure}[htbp]
    \centering
    \includegraphics[width=0.9\linewidth]{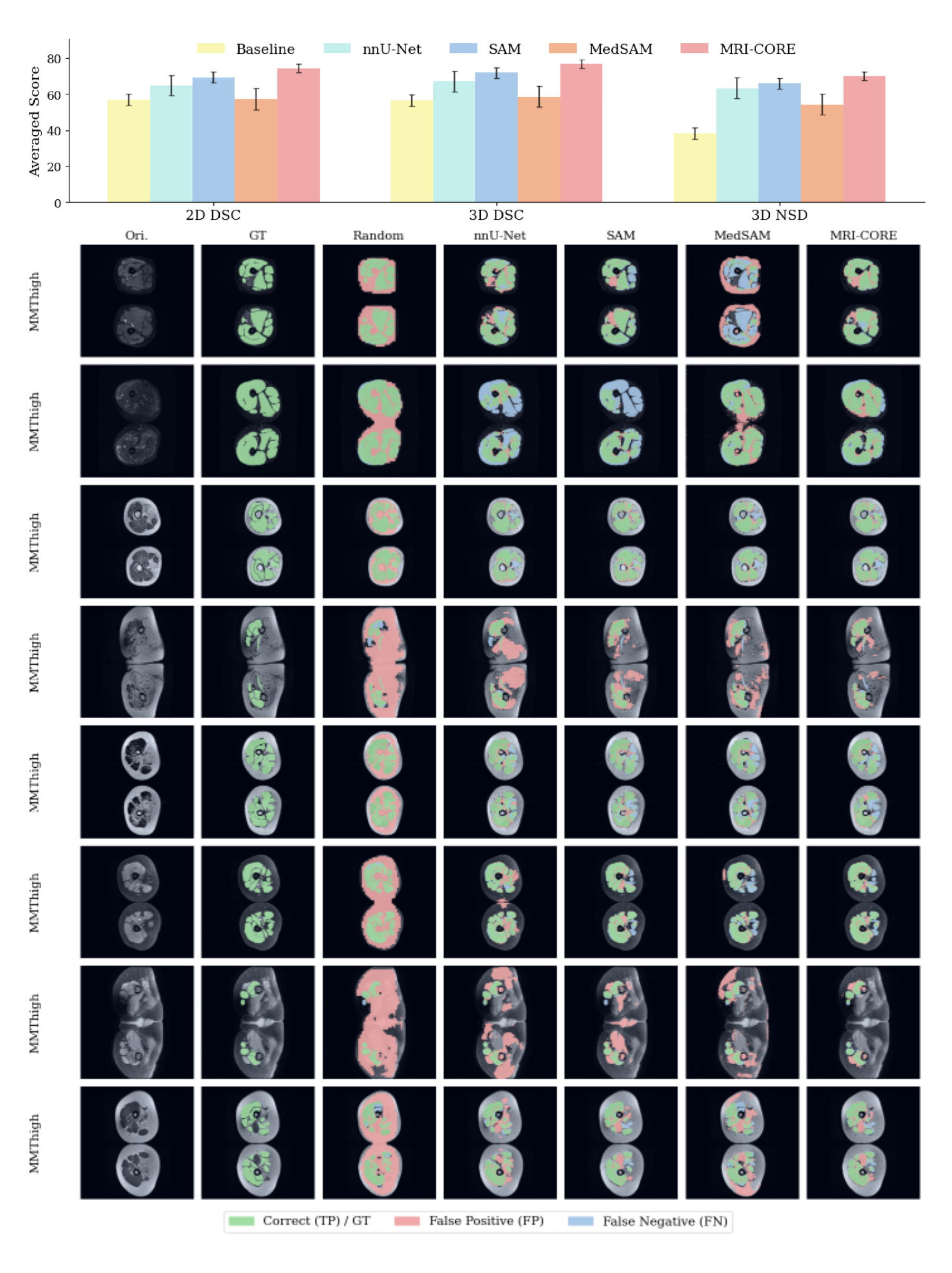}
    \caption{Qualitative results for few-shot segmentation on MMThigh dataset.
    }
    \label{fig:app_leg}
\end{figure}
\clearpage

\subsection{AMOS-Liver}
\begin{figure}[htbp]
    \centering
    \includegraphics[width=0.88\linewidth]{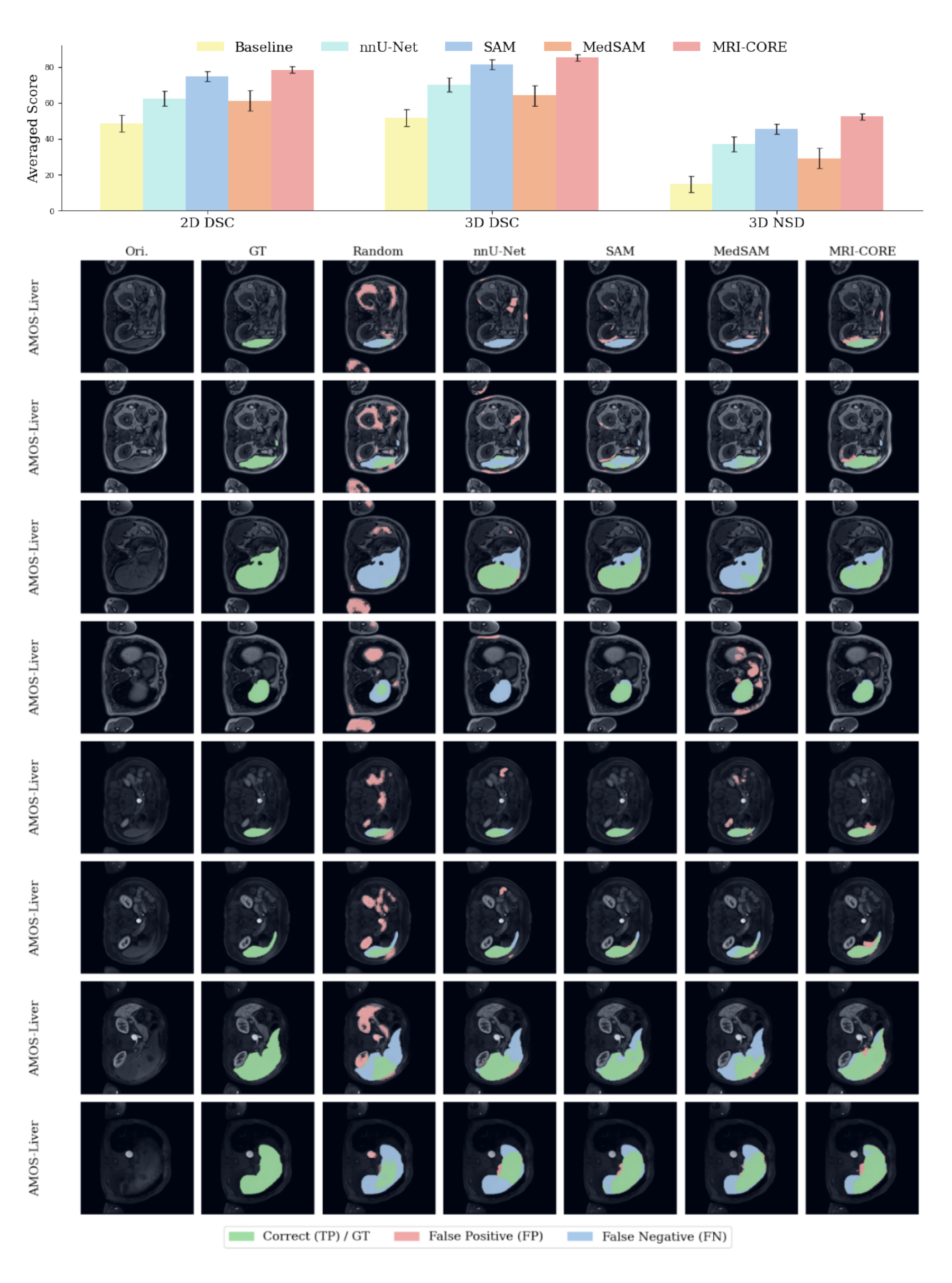}
    \caption{Qualitative results for few-shot segmentation on AMOS-Liver dataset.
    }
    \label{fig:app_liver}
\end{figure}
\clearpage

\subsection{AMOS-Pancreas}
\begin{figure}[htbp]
    \centering
    \includegraphics[width=0.88\linewidth]{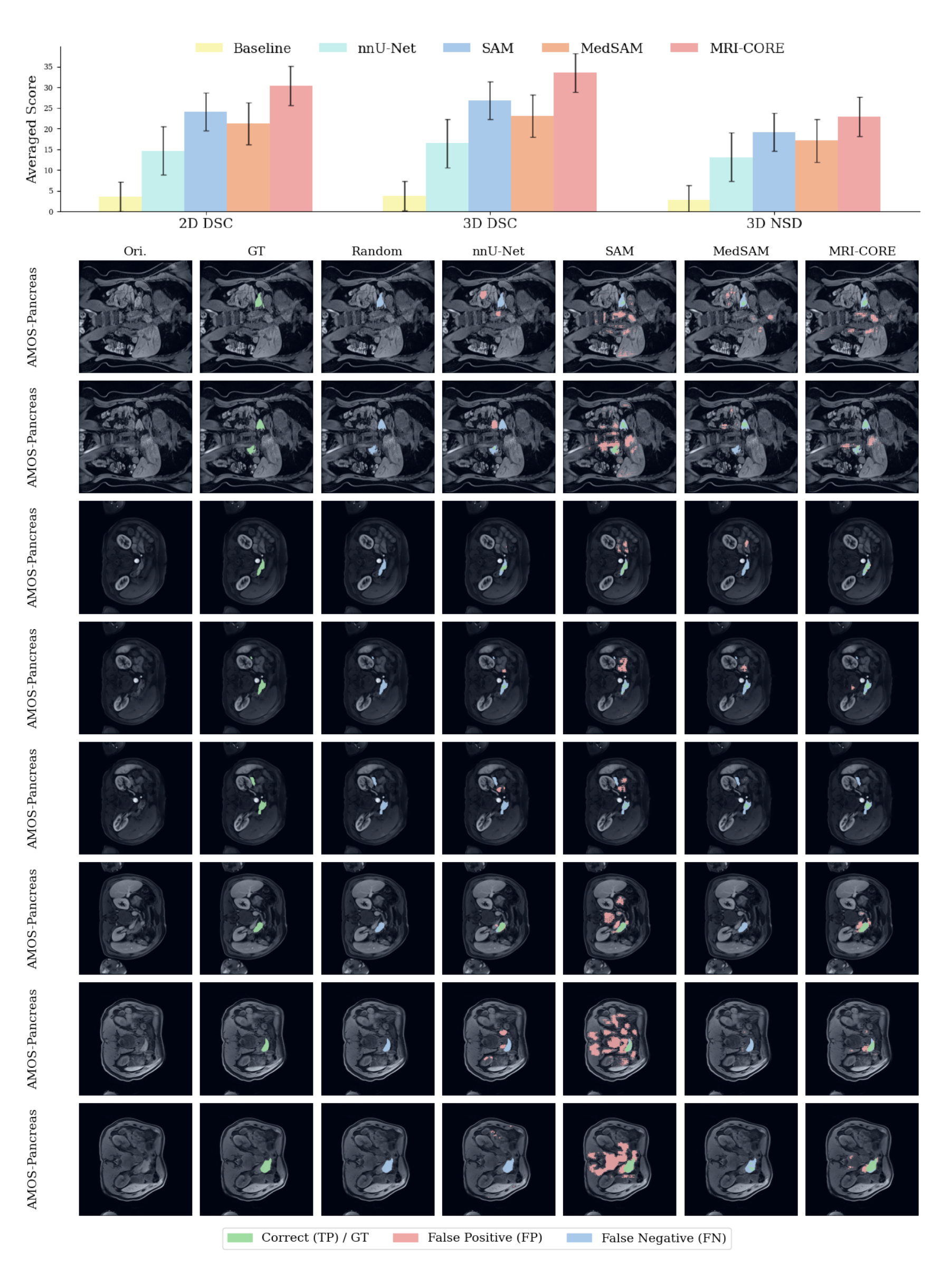}
    \caption{Qualitative results for few-shot segmentation on AMOS-Pancreas dataset.
    }
    \label{fig:app_pancreas}
\end{figure}
\clearpage

\subsection{AMOS-Spleen}
\begin{figure}[htbp]
    \centering
    \includegraphics[width=0.9\linewidth]{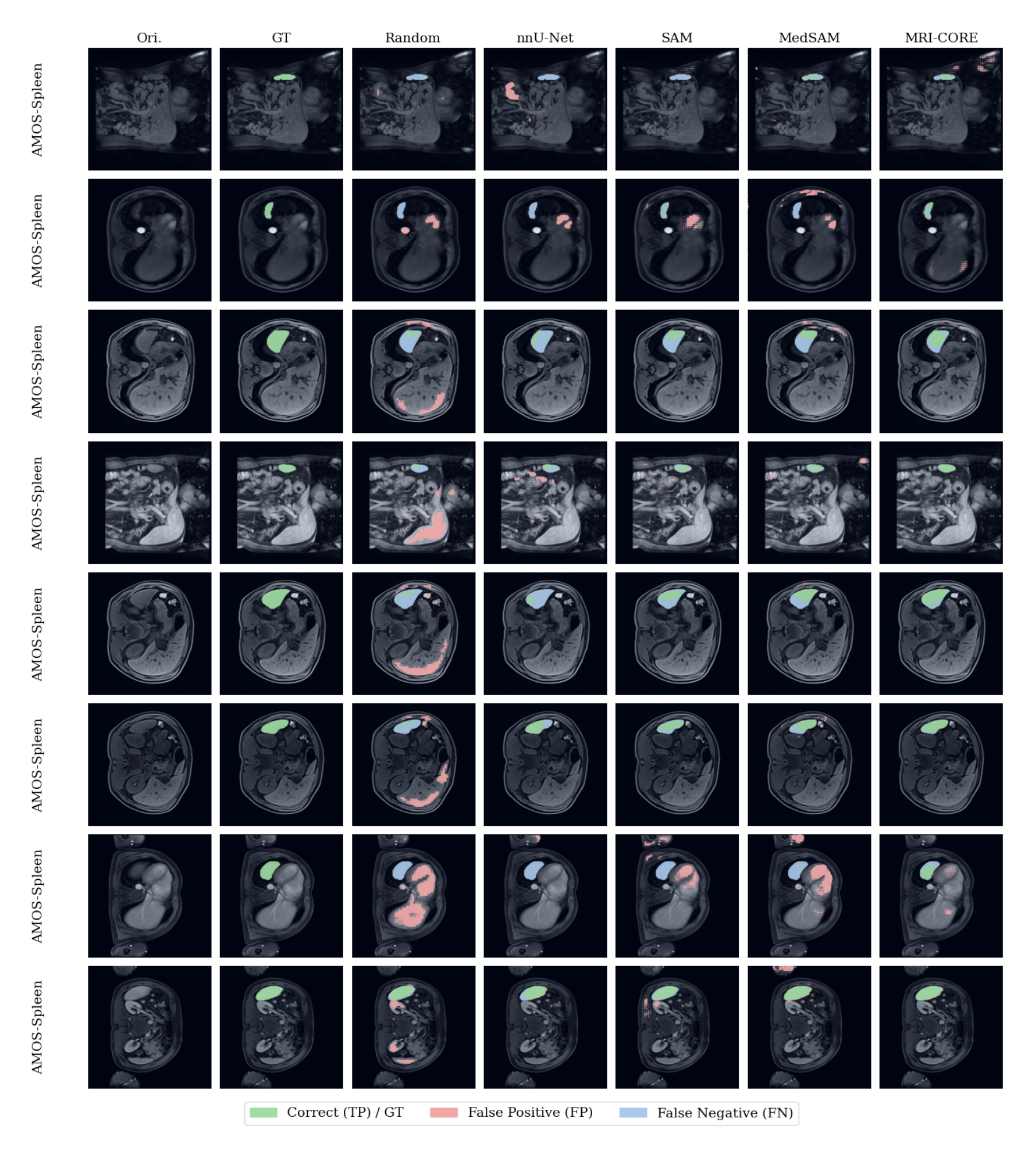}
    \caption{Qualitative results for few-shot segmentation on AMOS-Spleen dataset.
    }
    \label{fig:app_spleen}
\end{figure}
\clearpage

\subsection{SPIDER}
\begin{figure}[htbp]
    \centering
    \includegraphics[width=0.88\linewidth]{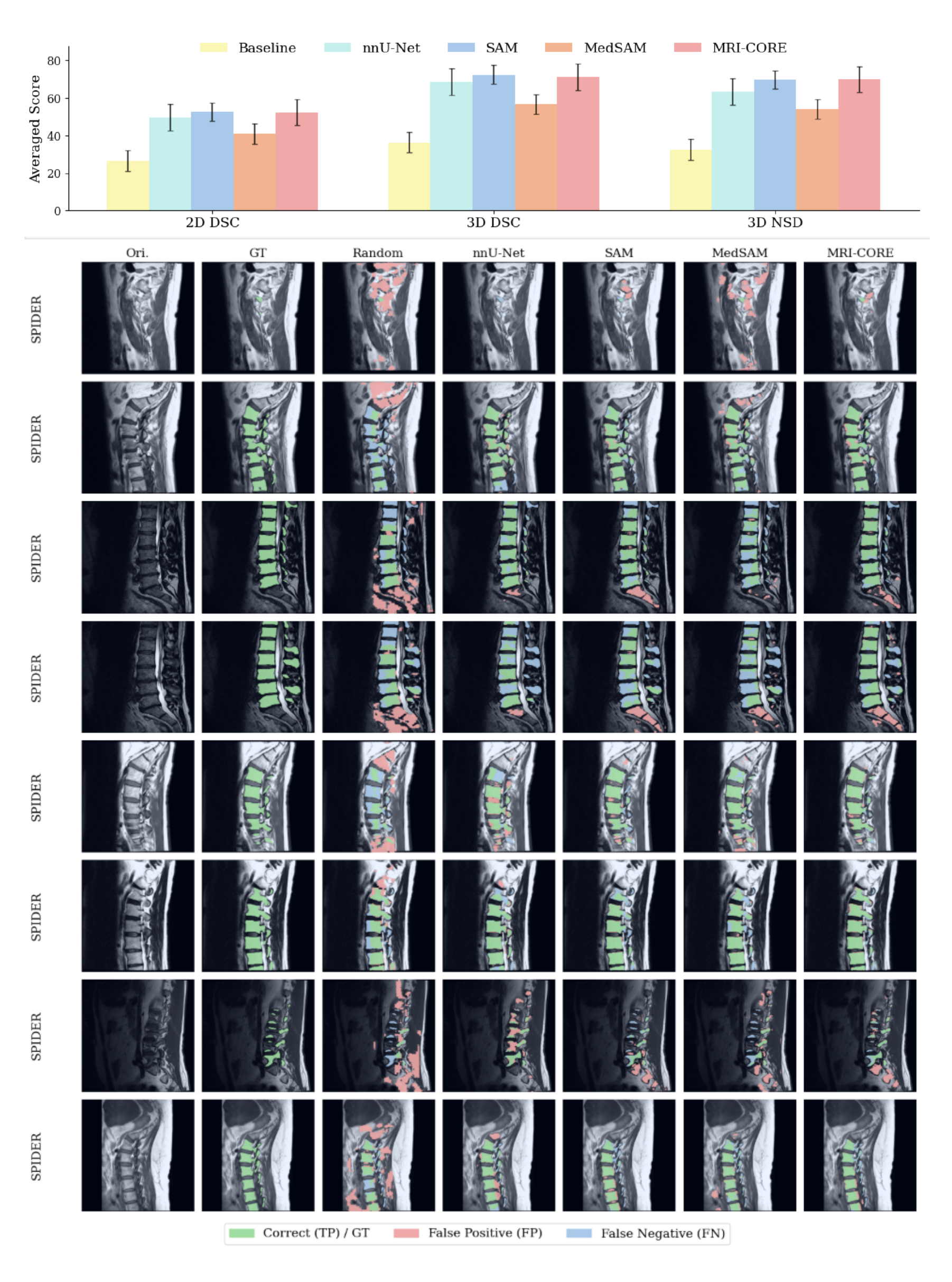}
    \caption{Qualitative results for few-shot segmentation on SPIDER dataset.
    }
    \label{fig:app_spider}
\end{figure}
\clearpage

\subsection{QIN \& ISBI-MR}
\begin{figure}[htbp]
    \centering
    \includegraphics[width=0.88\linewidth]{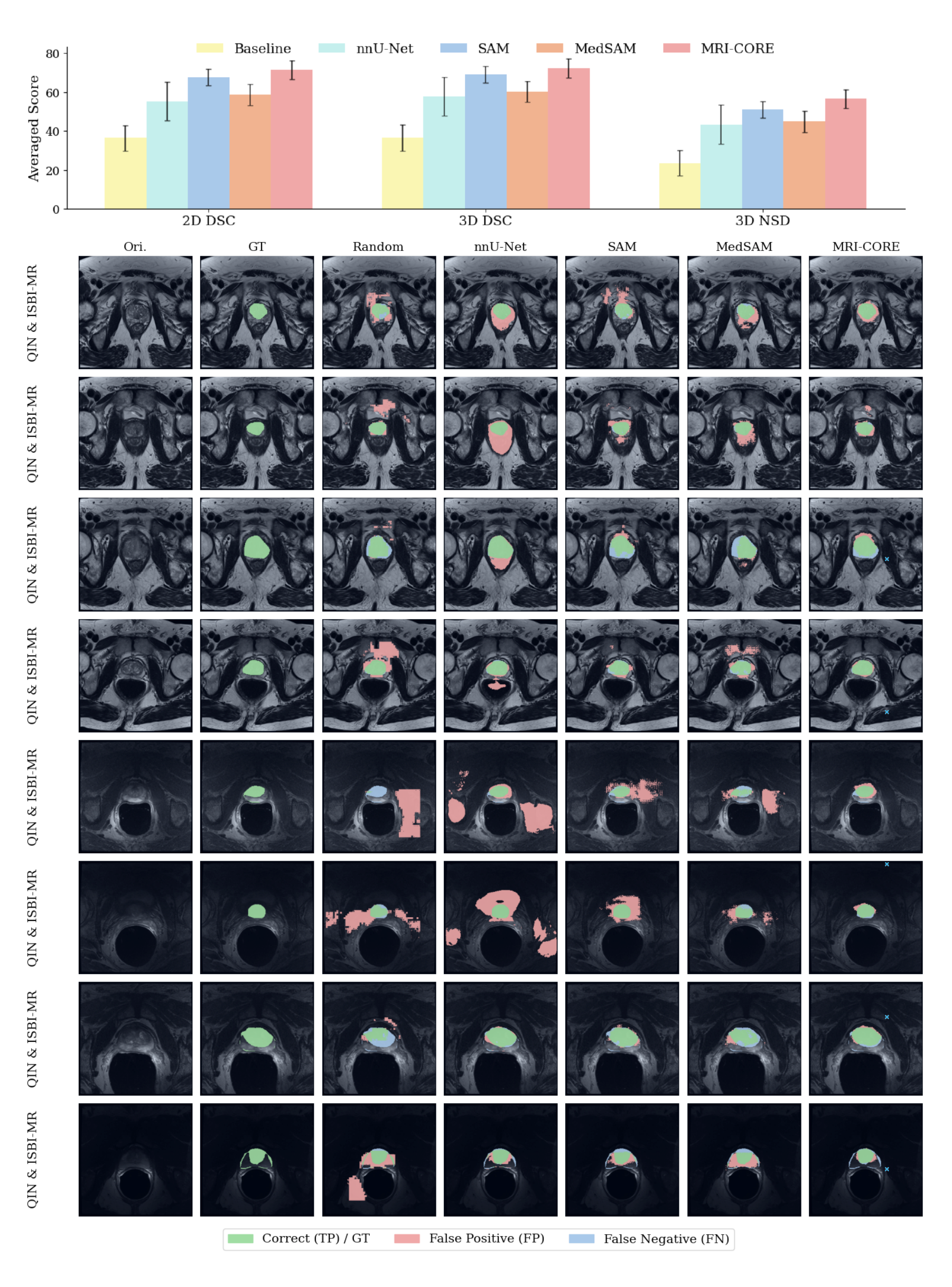}
    \caption{Qualitative results for few-shot segmentation on QIN \& ISBI-MR dataset.
    }
    \label{fig:app_prostate}
\end{figure}
\clearpage

\subsection{HVSMR-Chamber}
\begin{figure}[htbp]
    \centering
    \includegraphics[width=0.88\linewidth]{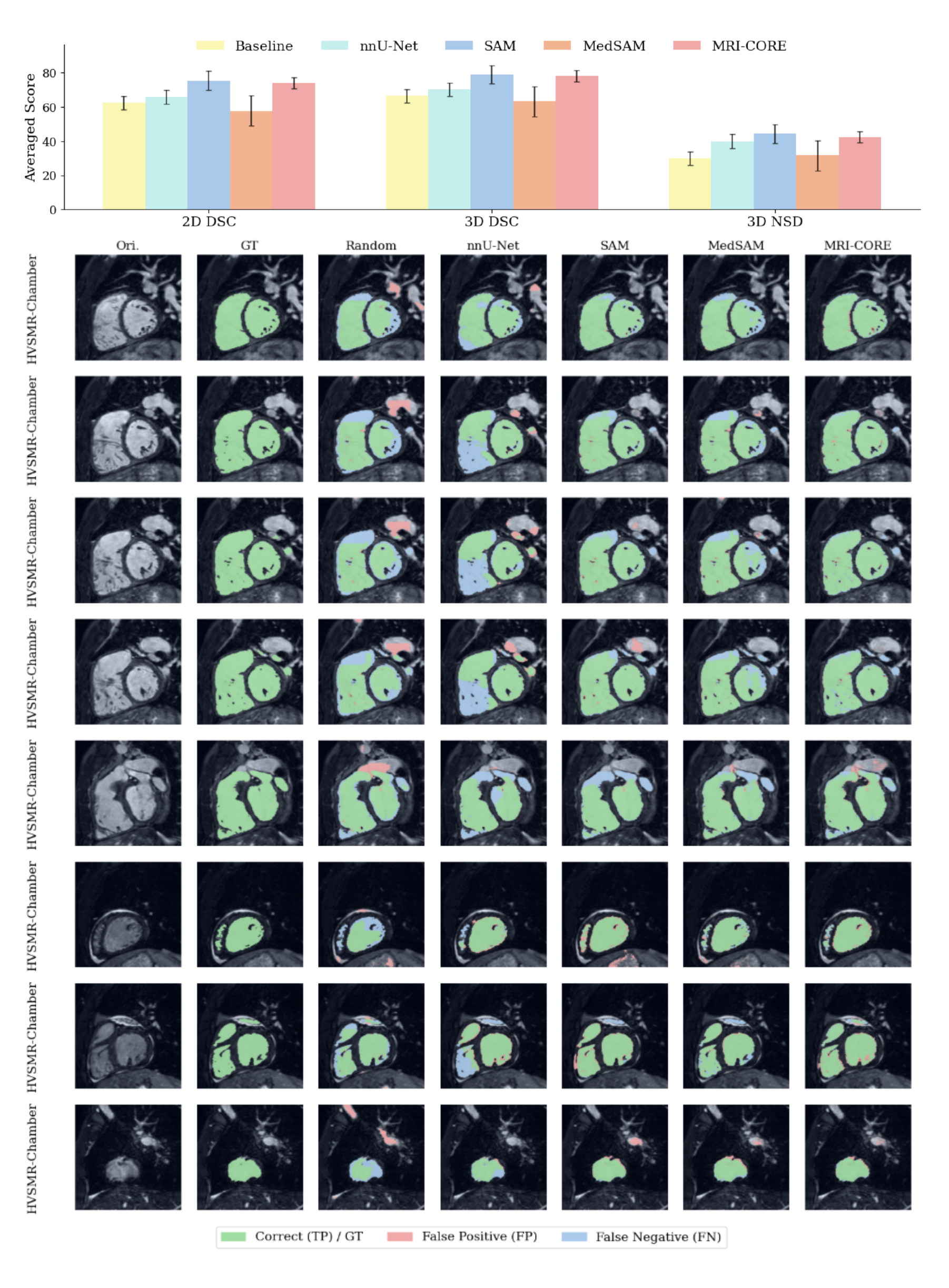}
    \caption{Qualitative results for few-shot segmentation on HVSMR-Chamber dataset.
    }
    \label{fig:app_chamber}
\end{figure}
\clearpage

\subsection{HVSMR-Vessel}
\begin{figure}[htbp]
    \centering
    \includegraphics[width=0.88\linewidth]{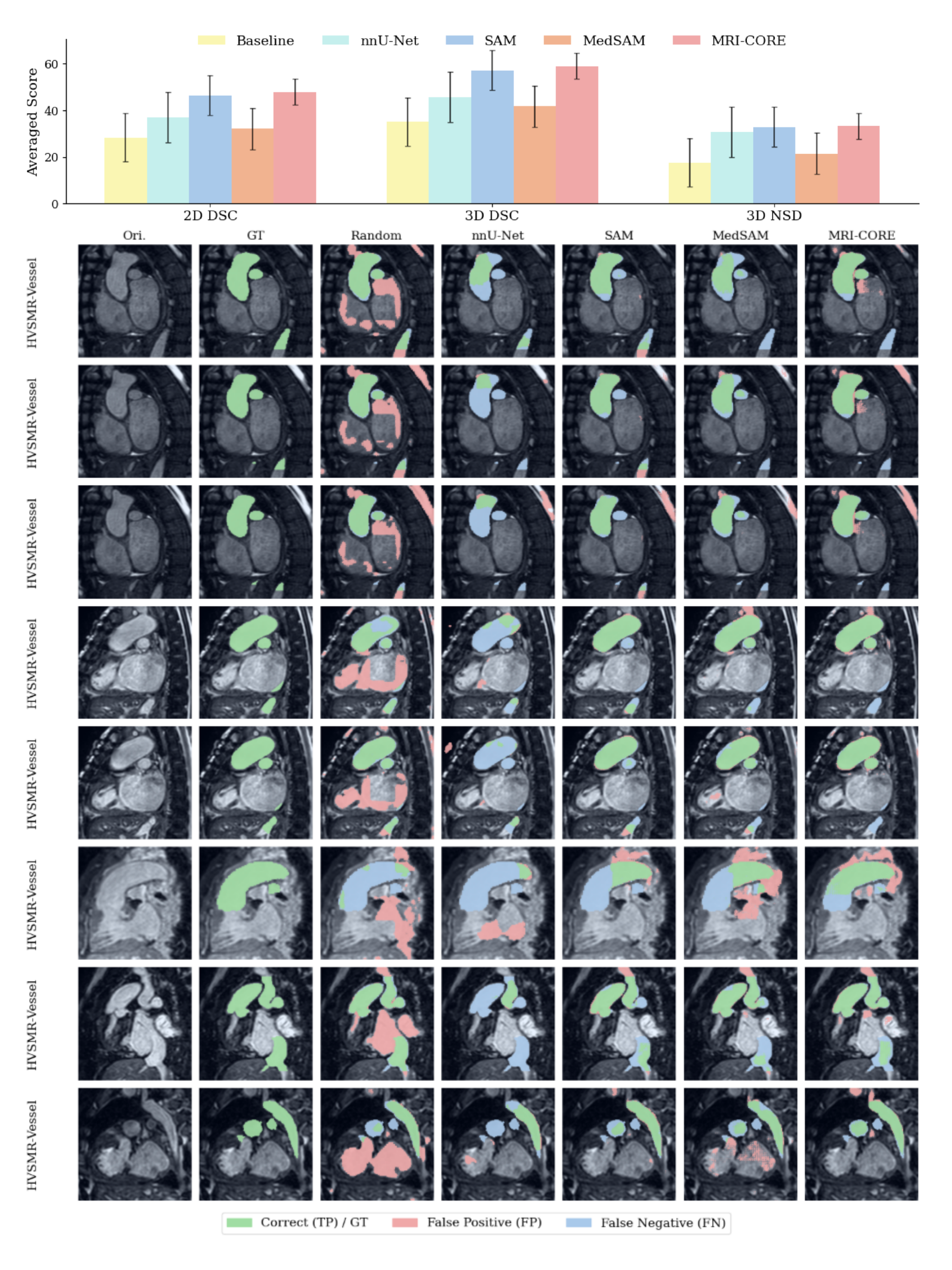}
    \caption{Qualitative results for few-shot segmentation on HVSMR-Vessel dataset.
    }
    \label{fig:app_vessel}
\end{figure}
\clearpage

\subsection{LiverHccSeg-Tumor}
\begin{figure}[htbp]
    \centering
    \includegraphics[width=0.88\linewidth]{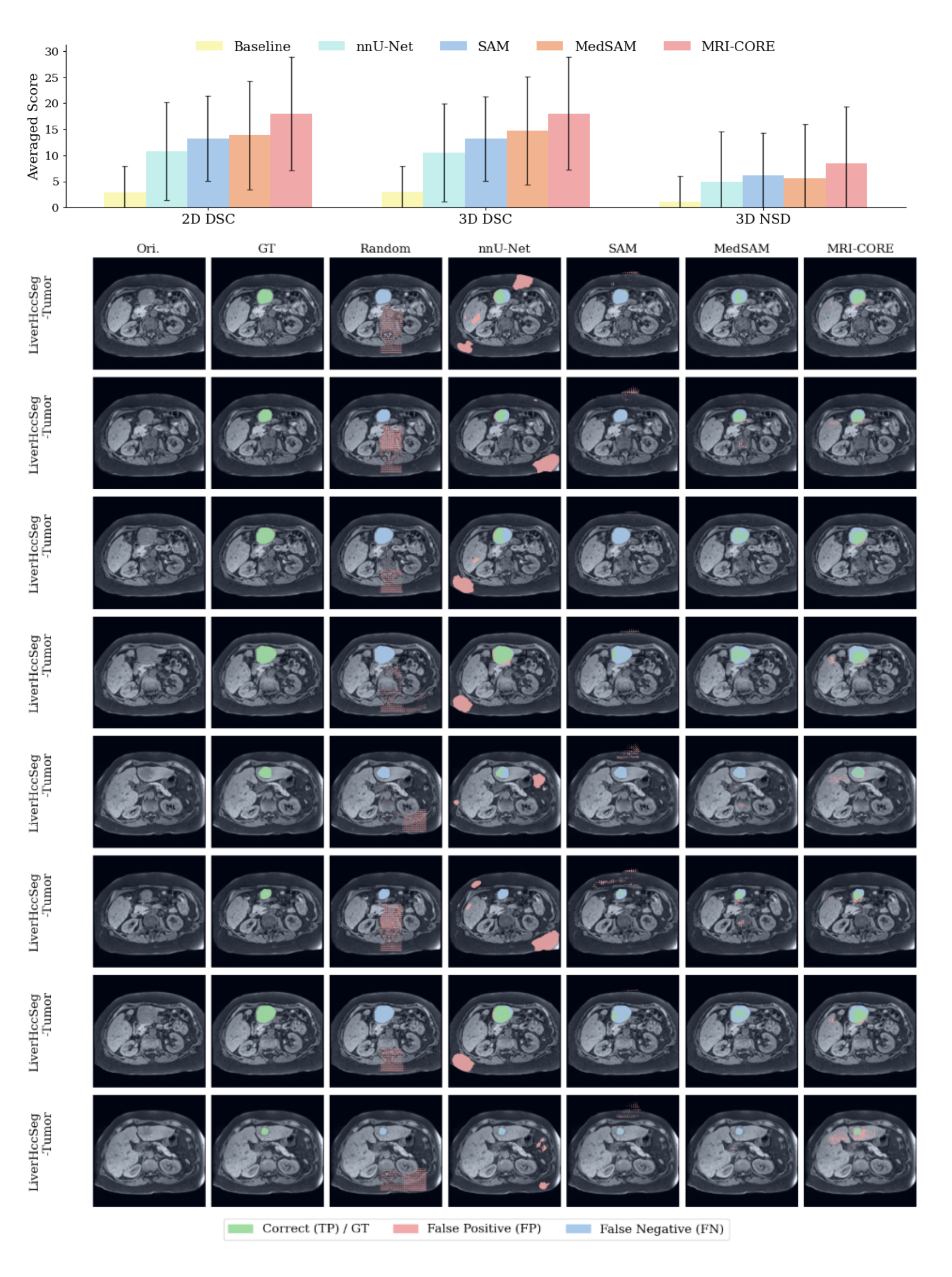}
    \caption{Qualitative results for few-shot segmentation on LiverHccSeg-Tumor dataset.
    }
    \label{fig:app_tumor}
\end{figure}
\clearpage

\end{appendices}

\end{document}